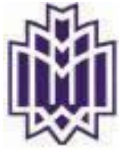

Tarbiat Moallem University
Humanities Faculty
Geography Department

A Dissertation for the Degree of PhD. In Physical Geography (Climatology)

Dissertation Title:
# The Analysis of Summertime Atmospheric Circulation over Iran and it's relation to Summertime Precipitation In Iran Plateau

Supervisors:
## Prof. B. Alijani     Dr. Z. Jafarpour

Advisor:
## Prof.  A.A. Bidokhti

By:
## A. Mofidi

November 2007

## Abstract:

In order to study the relation between the components of general circulation and summer precipitation over Iranian plateau, daily rainfall data of synoptic stations of Iran were obtained from IRIMO for the period 1970-2003. Surface and upper level data of NCEP for the same period were obtained from the CDC site of NOAA. The study area was defined as 20°E-120°E and 10°N-65°N.

Composite maps of different components such as pressure levels, vector wind, U and V winds, vertical velocity, Specific humidity, OLR, streamlines, vorticity, and temperature were drawn and analyzed.

The results showed that the summer circulation of atmosphere over South West Asia after a sudden change during late May to early June was established in its normal position. Different new pressure systems such as Zagross Trough were identified over the surface and Turkmenistan Anticyclone, the quasi-stationary trough of east Turkey at 700 hPa level. During the establishment of Subtropical High Pressure at the middle and upper troposphere over Iran, the summer circulation of the southwest Asia reached to its highest intensity and expansion, and continued to mid-August. On the other hand, the establishment of the Turkmenistan Anticyclone and Pakistan Low caused the 120 days winds of Sistan, and the development of a High pressure over Arabian Peninsula and Zagross Trough resulted in the Shomal wind of Khuzestan.

The main region of summer precipitation was identified as a triangular area to the east of 58.30°E and south of 28.30° N. Synoptic patterns responsible for the summer precipitation of the area are grouped into four categories as: Iran low pressure, Monsoon Depression, Pakistan low pressure, and local convective cells. Iran low pressure was responsible for more than 65 percent of summer rainfall, while the Monsoon and Pakistan lows produced only 15 percent of the rain. Most of the moisture of these rains (82%) came from the west of Oman sea and the northwest Arabian sea via the southwesterly flow below 850hPa level, whereas only 10 percent was from northeast of these water bodies. The moisture of Persian Gulf reached the area within the lower levels of the atmosphere.

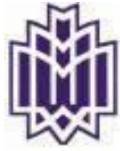

**دانشگاه تربیت معلّم**

**دانشکده ادبیات و علوم انسانی**

**گروه جغرافیا**

رساله جهت دریافت درجه دکتری تخصصی(PhD) در جغرافیای طبیعی-گرایش اقلیم شناسی

عنوان رساله:

# تحلیل گردش جو تابستانه بر روی ایران و ارتباط آن با بارش های تابستانه فلات ایران

اساتید راهنما:

**دکتر بهلول علیجانی    دکتر زین العابدین جعفرپور**

استاد مشاور:

**دکتر عباسعلی علی اکبری بیدختی**

دانشجو:

**عبّاس مفیدی**

آبان ماه ۱۳۸۶

## تشکر و قدر دانی:

پس از سپاس خداوند یکتا خداوند برخود فرض می دانم مراتب تشکر و کمال سپاسگزاری خود را تقدیم کسانی نمایم که در مراحل انجام این تحقیق مرا یاری داده اند.

از استاد دانشمند جناب آقای دکتر بهلول علیجانی بخاطر زحماتشان در به انجام رسیدن این رساله و همچنین به جهت لطف و عنایتی که در طول دوره تحصیل به اینجانب داشته اند تشکر فراوان دارم. از اساتید بزرگوار جناب آقای دکتر زین العابدین جعفرپور و جناب آقای دکتر عباسعلی علی اکبری بیدختی بخاطر پیشنهادات و بحث های سازنده شان بسیار سپاسگزارم. از داور محترم رساله، دانشمند فرزانه جناب آقای دکتر هوشنگ قائمی که در طی چند سال گذشته همواره با بحث های علمی مفید، رهنمودها و پیشنهادات سازنده شان اینجانب را در به انجام رسانیدن این تحقیق پشتیبانی و تشویق نموده اند بسیار سپاسگزارم. همچنین از سایر داوران محترم جناب آقای دکتر فرج زاده، جناب آقای دکتر سلیقه و سرکار خانم دکتر حجازی زاده که ضمن مطالعه رساله، نکات و پیشنهادات ارزشمندی را در جهت بهبود تحقیق یادآور شده اند تشکر فراوان دارم.

از جناب آقای مصطفی کریمی دانشجوی دکتری اقلیم شناسی دانشگاه تربیت مدرس، جناب آقای مهدی صداقت دانشجوی دکتری اقلیم شناسی دانشگاه تربیت معلم بخاطر راهنمائیهایشان و همچنین از جناب آقای محمد آخشیک به جهت تایپ بخشی از پیش نویس رساله کمال تشکر را دارم.

از خانم کتی اسمیت بخاطر راهنمایی های مکررشان در استفاده از پایگاه اینترنتی NCEP و از دکتر چارلز سمن به جهت تصحیح برخی از اسکریپت های مربوط به نرم افزار GrADS و همچنین از دکتر غلام رسول از دپارتمان هواشناسی پاکستان بخاطر در اختیار قرار دادن داده های بارش روزانه ایستگاه های پاکستان بسیار متشکرم. از سرکار خانم آذر زرین به جهت پیشنهادات و بحث های مفیدشان، همچنین تایپ بخشی از پیش نویس رساله، ویرایش متن و هم فکری در تهیه برخی اسکریپت ها و از سرکار خانم موسوی نماینده تحصیلات تکمیلی دانشکده ادبیات و علوم انسانی دانشگاه تربیت معلم به جهت مساعدت فراوان درتسهیل امور بسیار سپاسگزارم.

**چکیده:**

به منظور تبیین نحوه ارتباط بین مؤلفه های اصلی گردش جو بر روی جنوب غرب آسیا با بارش های تابستانه بخش های داخلی فلات ایران و تعیین الگوهای سینوپتیکی حاکم بر بارش ها، از داده های بارش روزانه ایستگاه های سینوپتیک کشور و داده های رقومی مراکز ملی پیش بینی محیطی آمریکا(NCEP/NCAR) برای یک دوره ۳۴ ساله(۲۰۰۳-۱۹۷۰) استفاده شد.

با استفاده از داده های رقومی ارتفاع ژئوپتانسیل، مؤلفه های باد مداری و باد نصف النهاری، سرعت قائم، نم ویژه، دما و تابش موج بلند خروجی(OLR) برای مقیاس های زمانی و ترازهای مختلف، نقشه های ترکیبی ارتفاع ژئوپتانسیل، تاوائی نسبی، خطوط جریان، باد بُرداری، دیاگرام هاومولر(نمودارx-t)، و همچنین مقاطع قائم از تاوائی نسبی، سرعت قائم و باد مداری و باد نصف النهاری تولید و مورد تحلیل قرار گرفت.

بررسی ساختار گردش جو تابستانه نشان دهنده ی آن است که گردش جو بر روی جنوب غرب آسیا در حد فاصل زمانی پایان ماه مه تا پایان دهه اول ماه ژوئن ضمن یک تغییر نسبتاً ناگهانی، در ساختار و موقعیت تابستانه خود ظاهر می گردد. از نتایج مهم پژوهش حاضر شناسایی واچرخند ترکمنستان، ناوه شبه ساکن شرق ترکیه و زبانه ی کم فشار زاگرس و همچنین درک ویژگی های سینوپتیکی و دینامیکی حاکم بر مؤلفه های مذکور و سایر مؤلفه های گردشی حاکم در ترازهای زیرین جو شامل کم فشار پاکستان و پرفشار عربستان بر روی منطقه جنوب غرب آسیا است. بررسی روند تکوین گردش جو تابستانه نشان داد که در روزهای آغازین ماه ژوئیه، در پی ظهور پرفشار جنب حاره ایران در ترازهای میانی و فوقانی وردسپهر، گردش جو تابستانه بر روی جنوب غرب آسیا به حداکثر شدت و گستردگی خود رسیده و این وضعیت تا پایان دهه دوم ماه اوت تداوم می یابد. همچنین یافته ها نشان دهنده آن است که شکل گیری و تداوم باد۱۲۰ روزه سیستان ناشی از ظهور و تداوم واچرخند ترکمنستان و کم فشار پاکستان و شیب فشار ناشی از استقرار مراکز فشار مذکور در ترازهای زیرین جو بر روی نیمه شرقی فلات ایران می باشد. در مقابل تشکیل و تداوم باد شمال نیز نتیجه استقرار مرکز پرفشار عربستان در موقعیت نرمال تابستانه خود همراه با شکل گیری و تداوم زبانه ی کم فشار زاگرس بر روی نیمه غربی فلات ایران است.

نتایج بررسی ها نشان داد که بارش های تابستانه‌ی ایران در دو منطقه‌ی اصلی شامل نوار شمالی با مرکزیت سواحل جنوبی دریای خزر و منطقه‌ی جنوب شرق کشور فرو می‌ریزد. نقشه‌ی بارش فصلی همچنین هسته‌های بارشی کوچکتری را بر روی جنوب استان فارس و جنوب غرب استان کرمان نشان داد. بررسی بارش های تابستانه مناطق داخلی فلات ایران بیانگر آن است که منطقه‌ی اصلی بارش، محدوده‌ی مثلثی شکلی در شرق ˚۵۸/۳۰ طول شرقی و جنوب عرض ˚۲۸/۳۰ شمالی درجنوب شرق کشور است که رأس آن را ایستگاه ایرندگان و قاعده‌ی آن را ایستگاه‌های چانف، اسپاکه و آشار با مقادیر متوسط بارش تابستانه بین ۴۵ تا ۶۵ میلی‌متر تشکیل می‌دهند. یافته ها نشان داد که شاخص منطقه‌ای جنوب شرق، معرّف اصلی بارش‌های تابستانه‌ی فلات ایران به شمار رفته و تغییرات بارشی منطقه‌ی شاخص، بیانگر تغییرات بارش مناطق داخلی ایران است. بر این اساس ویژگی های غالبِ بارش های تابستانه فلات ایران از منظر مکانی و زمانی در مقیاس منطقه ای مورد بررسی قرار گرفت.

نتایج بررسی۳۹۲ روز در قالب ۵۶ دوره‌ی بارشی اولیه منجر به استخراج تعداد ۴ الگوی سینوپتیکی اصلی شامل: (۱)کم‌فشار ایران، (۲)کم‌فشار موسمی هند، (۳)کم‌فشار پاکستان و (۴) همگرایی‌های کم‌عمق، در قالب ۶۶ دوره بارشی نهائی برای بارش‌های تابستانه مناطق جنوب و جنوب شرقی ایران گردید. تحلیل جریان هوا در ترازهای زیرین و میانی وردسپهر نشان داد که علت اصلی وقوع بارش های تابستانه در مناطق داخلی فلات ایران، تشکیل و گسترش مرکز کم فشار کم عمقی موسوم به «کم فشار ایران» می باشد. این کم فشار در ۶۵٪ دوره های بارشی سازوکار اصلی وقوع بارش های تابستانه ایران بوده و در۱۰٪ از موارد دیگر نیز در وقوع بارش ها مشارکت داشته است. یافته ها بیانگر آن است که کم فشار موسمی هند و بطور کلی بارش های مرتبط با سیستم گردش موسمی تابستانه جنوب آسیا همچون کم فشار پاکستان تنها ۱۵٪ از بارش های تابستانه مناطق داخلی فلات ایران را موجب گردیده است.

نتایج بررسی ها نشاندهنده آن است که عمده رطوبت بارش های تابستانه مناطق داخلی فلات ایران (۸۲٪) در زیر تراز ۸۵۰ هکتوپاسکال از نیمه غربی دریای عمان و شمال غرب دریای عرب تأمین می گردد. همچنین یافته ها نشان داد که تنها در حدود ۱۰٪ از موارد، بواسطه تقویت و گسترش کم فشارهای موسمی هند، رطوبت در ترازهای میانی از قطاع شمال شرقی دریای عرب و غرب هند به منطقه جنوب-جنوب شرق ایران انتقال می یابد. نتایج بررسی ها همچنین بیانگر آن است که در زمان وقوع بارش های تابستانه در ایستگاه‌های واقع در جنوب کشور (از شمال تنگه هرمز به سمت غرب)، خلیج فارس نیز در پائین ترین ترازهای جو منبع رطوبت بارش ها بوده است.

**واژگان کلیدی:** گردش جو تابستانه، بارش تابستانه ایران، پرفشار جنب حاره، واچرخند ترکمنستان، ناوه ی شبه ساکن شرق ترکیه، کم فشار ایران، تحلیل جریان هوا.

ج

# فصل سوّم

## ساختار گردش جوّ تابستانه بر روی جنوب غرب آسیا

## 3-1. مقدمه

نحوه‌ی استقرار مراکز فشار در جو زمین، ضمن ایجاد و کنترل مستقیم گردش بزرگ مقیاس جو توزیع انرژی و مبادله‌ی آن را بر روی زمین کنترل نموده و تفاوت‌های موجود در توزیع عناصر جوی از جمله بارش را در سطح زمین به دنبال دارد. بر این اساس، جهت تبیین اقلیم تابستانه جنوب غرب آسیا و سازوکار حاکم بر وقوع بارش‌های تابستانه‌ی فلات ایران، خصوصیات مراکز فشار و ساختار گردش هوا بعنوان عامل بلافصل کنترل کننده‌ی آن از اهمیت خاصی برخوردار است. در ادامه ساختار گردش جو تابستانه بر روی جنوب و جنوب غرب آسیا مورد بررسی قرار می‌گیرد.

## 3-2. گردش بزرگ مقیاس جو بر روی جنوب و جنوب غرب آسیا

با آغاز دوره‌ی گرم سال واکنش گردش بزرگ مقیاس به گرمایش متفاوت بین توده‌ی خشکی قاره‌ی آسیا و اقیانوس هند در جنوب آن، توزیع انرژی و مبادلات حرارتی را در منطقه‌ی جنوب آسیا دستخوش تحول اساسی می‌نماید. شار گرمای محسوس بر روی فلات مرتفع تبت به همراه آزاد شدن گرمای نهان در شمال هند واداشت‌های حرارتی لازم را جهت تشکیل و تکوین یک سیستم گردش بزرگ مقیاس فصلی در منطقه‌ی جنوب و جنوب غرب آسیا فراهم می‌نماید (Staff Members, 1958; Flohn, 1957; Li *et al.*, 2001). در این دوره بواسطه جذب مقادیر قابل ملاحظه‌ی انرژی خورشیدی و وجود حجم قابل ملاحظه‌ی منابع گرمایش دیاباتیک (دررو) و آدیاباتیک (بی دررو)، دما در شمال هند و بر فراز فلات تبت از دمای میانگین جو آزاد در همان ارتفاع بیشتر می‌شود (شکل3-1a). بطوریکه ناهنجاری دما در سطح فلات تبت به بیش از 4 درجه‌ی سانتیگراد و در تراز 200 هکتوپاسکال به حداکثر میزان خود یعنی حدود 6/5 درجه‌ی سانتیگراد می‌رسد (Yeh, 1981; 1982). این ناهنجاری بالاترین ناهنجاری دمایی در کل کره‌ی زمین در این ارتفاع محسوب می‌گردد (شکل3-1a). در واقع فلات تبت بعنوان «یک منبع گرمای محسوس مرتفع» در وردسپهر میانی عمل می‌کند و به خاطر ارتفاع زیاد فلات، گرما مستقیماً به وردسپهر میانی اضافه شده و موجب گرمایش نیمه‌ی فوقانی جرم جوی می‌شود (Yeh, 1981; 1982). شکل‌گیری بزرگترین «چشمه‌ی گرمایی» در شمال هند و بر فراز فلات تبت با رعایت تقدم زمانی، تشکیل و یا تقویت مؤلفه‌های زیر را بعنوان مؤلفه‌های اصلی گردش جو تابستانه در منطقه‌ی آسیا در پی خواهد داشت:

1- گرمایش نیمه‌ی فوقانی وردسپهر بر روی فلات، موجب پیدایش گردش چرخندی و کم‌فشار حرارتی در ترازهای زیرین جو و گردش واچرخندی و پرفشار هسته‌ی گرم در وردسپهر فوقانی می‌گردد (Flohn, 1957; Krishnamurti *et al.*, 1973a). این پرفشار هسته‌ی گرم، قوی‌ترین و در عین حال گسترده‌ترین پرفشار لایه‌های فوقانی وردسپهر در کره‌ی زمین محسوب می‌گردد (Mason and Anderson, 1963).

2- گرمایش نیمه‌ی فوقانی وردسپهر بر روی فلات تبت ضمن ایجاد پرفشار هسته‌ی گرم در جنوب آسیا، شیب دما (شکل1a-3) و فشار را در قطاع آسیای جنوبی- اقیانوس هند معکوس می‌نماید که همین امر موجب تضعیف شدید گردش هدلی و معکوس شدن گردش نصف النهاری و آغاز موسمی جنوب شرق آسیا در فصل بهار نیمکره‌ی شمالی می‌گردد (Flohn, 1957; Koteswaram, 1958).

3- مناطق خشک و بیابانی جنوب غرب آسیا، سرمایش تابشی خالصی را بویژه در وردسپهر فوقانی به رغم گرمایش شدید تراز زیرین جو در تابستان تجربه می‌کنند. از نظر تئوریکی سرمایش ایجاد شده بایستی بواسطه‌ی فرارفت قائم به توازن برسد(Webster, 1983). بر این اساس، فرونشینی هوا بر روی مناطق خشکِ وسیع و بدون ابرِ جنوب غرب آسیا جهت ایجاد توازن در سرمایش تابشی ضروری است ( ,.Webster et al 1998). با توجه به شکل2a-3 در دوره‌ی گرم سال، مناطق بیابانی جنوب غرب آسیا بواسطه‌ی خروج قابل ملاحظه‌ی انرژی از زمین، همچون بسیاری از مناطق خشک و بیابانی دنیا بصورت یک «چاهه‌ی گرمایی» جلوه‌گر می‌شود(Charney, 1975)، که نتیجه‌ی مستقیم آن پیدایش یک شیب گرمایی قابل ملاحظه در امتداد طولی(مداری) بین منطقه‌ی چشمه(فلات تبت) و چاهه‌ی گرمایی(جنوب غرب آسیا)، (شکل2a-3 و2b-3) و شکل‌گیری گردش بزرگ مقیاس مداری، بواسطه‌ی صعود در منطقه‌ی چشمه‌ی انرژی و نزول دینامیکی در منطقه‌ی چاهه‌ی جنوب غرب آسیاست ( ,.He et al., 1987;Yanai et al., 1992;Webster et al 1998). مطالعات متعدد وجود چنین گردشی را بر روی صفحه‌ی مداری تأیید نموده‌اند و نام‌هایی چون«گردش شرقی-غربی تابستانی» یا «گردش موسمی متقاطع»[32] را برای آن برگزیده‌اند (Krishnamurti,1971a;Yang et al.,1992; Webster,1994; Webster et al., 1998; Meehl, 2003). شایان ذکر است که چشمه‌ی گرمایی جنوب آسیا عامل به حرکت درآورنده‌ی چنین گردشی محسوب می‌گردد.

4- تداوم نزول آدیاباتیک در غرب سوی فلات تبت ضمن تشکیل و تقویت مرکز پرفشار بر روی ایران، جابجایی شمال سوی ناگهانی جت جنب حاره را بر روی مناطق جنوب و جنوب غرب آسیا بدنبال دارد. در عین حال گرمایشِ ناشی از نزول آدیاباتیک همراه با گرمایش محسوس در فلات غربی و جنوب غرب آسیا معکوس شدن شیب دما و فشار را در منطقه‌ی غرب فلات تبت در امتداد نصف النهاری در پی داردکه نتیجه‌ی آن «آغازگری»[33] موسمی تابستانه بر روی شبه قاره‌ی هند و دریای عرب می‌باشد ( He et al., 1987; Yanai et al., 1992).

---

32.Transverse Monsoon

33.Onset

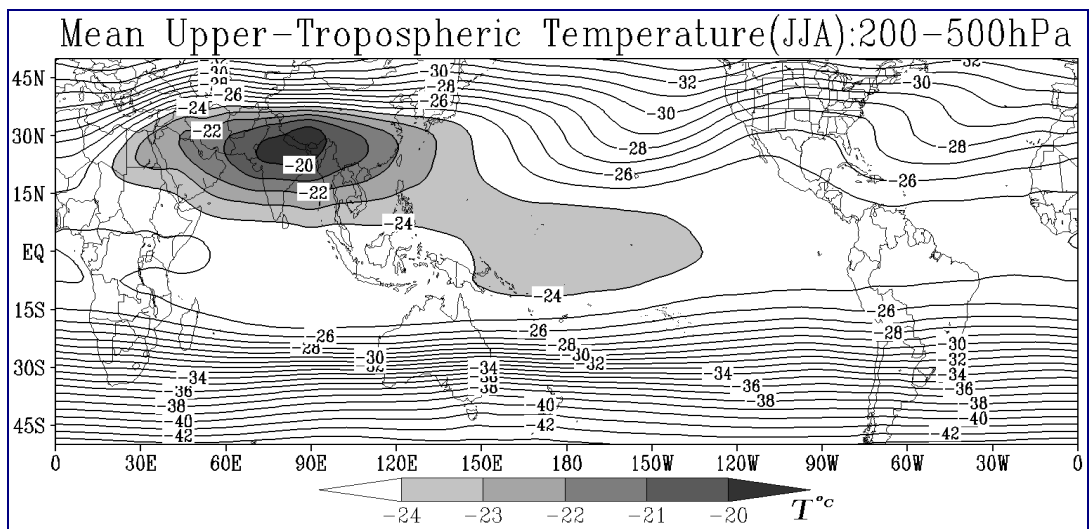

**شکل۱-۳.** دمای متوسط درازمدت فصل تابستان در نیمه فوقانی وردسپهر متوسط گیری شده برای ترازهای ۵۰۰ تا ۲۰۰ هکتوپاسکال. دمای گرم تر از ۲۵°- سانتیگراد بصورت سایه دار به نمایش در آمده است. جهت محاسبه از داده های یک دوره ۲۹ ساله(۱۹۹۶-۱۹۶۸) ماه های ژوئن، ژوئیه و اوت استفاده شده است.

۵- با آغازگری موسمی جنوب آسیا، بواسطه‌ی آزاد شدن حجم قابل ملاحظه‌ی گرمای نهان تبخیر، چشمه‌ی گرمایی در شمال هند مجدداً تقویت شده که نتیجه‌ی مستقیم آن تقویت و گسترش پرفشار هسته‌ی گرم در جنوب آسیا و بدنبال آن تقویت گردش هدلی موسمی در امتداد نصف النهاری و گردش مداری در امتداد صفحه‌ی مداری در آغاز تابستانی نیمکره‌ی شمالی است. در این زمان جت شرقی تراز فوقانی و جت غربی تراز زیرین در غرب اقیانوس هند و بر روی دریای عرب استقرار می یابند (Krishnamurti, 1985).

بر این اساس، در بررسی ساختار قائم گردش جو تابستانه، در سرتاسر خشکی‌های جنوب و جنوب غرب آسیا، در ترازهای زیرین وردسپهر شاهد حضور مراکز همگرایی و کم‌فشار و در ترازهای فوقانی شاهد استقرار مراکز واگرایی و پرفشار هستیم (Krishnamurti, 1971b; Wu and Liu, 2003; Chen, 2003; Chen et al., 2005).

با توجه به مطالب فوق، پرفشارها و جت جنب حاره در تراز فوقانی و مراکز کم‌فشار کم‌عمق در تراز زیرین ویژگی‌های اقلیمی غالب منطقه‌ی جنوب غرب آسیا در فصل تابستان به شمار می‌روند. در ادامه تشکیل و تکوین مؤلفه‌های اصلی گردش جو تابستانه بر روی جنوب غرب آسیا مورد بررسی قرار می‌گیرد.

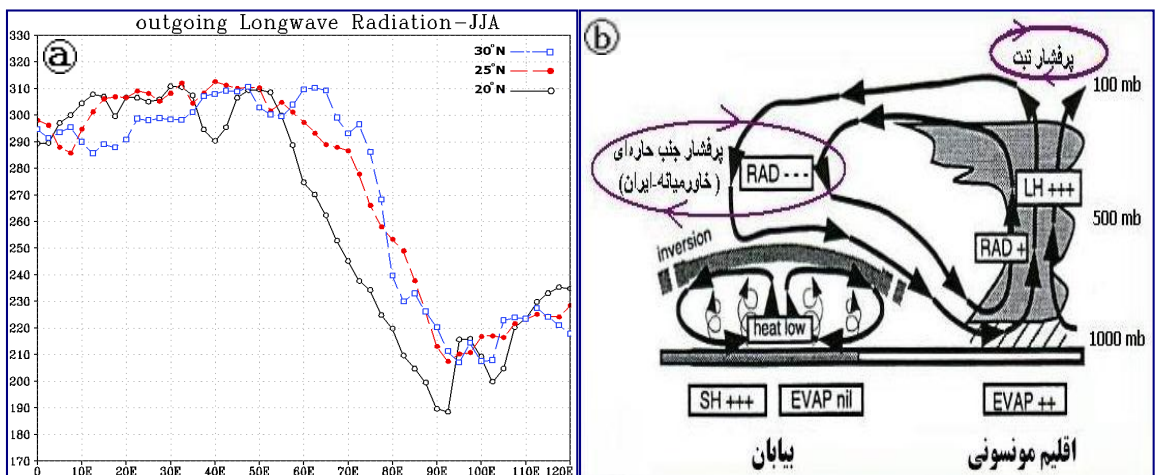

**شکل۳-۲.** (a) مقادیر متوسط درازمدت OLR در امتداد عرض های ۲۰°، ۲۵° و ۳۰° شمالی برای فصل تابستان برحسب وات بر مترمربع. جهت محاسبه از داده های یک دوره ۲۹ ساله(۱۹۹۶-۱۹۶۸) ماه های ژوئن، ژوئیه و اوت استفاده شده است. (b) ارتباط واداشت های حرارتی در جنوب آسیا و گردش مداری غرب سو بر روی جنوب غرب آسیا. مستطیل ها میزان گرمایش یا سرمایش مسلط را نشان می دهند. حروف SH، LH، RAD و EVAP به ترتیب گرمایش محسوس، نهان، تابشی و تبخیر را بیان می نمایند(Webster et al., 1998).

## ۳-۲-۱. گردش جو تابستانه در ترازهای میانی و فوقانی
### ۳-۲-۱-۱. پرفشارهای جنب حاره
#### الف- پرفشار جنب حاره تابستانه در تراز فوقانی

در طول ماه‌های تابستانی واچرخند بزرگی بر روی مناطق جنوب و جنوب غرب آسیا استقرار می‌یابد که در تراز ۲۰۰ هکتوپاسکال به حداکثر شدت و گستردگی خود می‌رسد. رفتار این واچرخند توسط پژوهشگران متعددی مورد بررسی قرار گرفته است(Flohn, 1957; Krishnamurti, 1971a; 1973a; Reiter et al., 1982; Zhang et al., 2002).

به منظور شناسایی موقعیت مکانی و وردش‌های زمانی پرفشار جنب حاره در تراز فوقانی، مقادیر ارتفاع ژئوپتانسیل و تاوایی نسبی در تراز ۲۰۰ هکتوپاسکال برای یک دوره‌ی اقلیمی ۲۹ ساله بصورت میانگین پنج روزه از ابتدای آوریل تا انتهای اکتبر مورد بررسی قرار گرفت. شکل۳-۳ وردش‌های زمانی و مکانی پرفشار جنب حاره را بر روی جنوب و جنوب غرب آسیا نشان می‌دهد. بررسی نقشه‌ها بیانگر آن است که پرفشار جنب حاره در طی روزهای پایانی ماه آوریل بر شمال برونئی شکل می‌گیرد و در طی پنج روز اول ماه می بر روی ویتنام جنوبی و کامبوج در عرض۱۲° استقرار می‌یابد (شکل۳a-۳). مرکز واچرخند تا پایان ماه می در یک جابجایی شمال غرب سو به بالای عرض ۲۰° شمالی منتقل شده و بر روی شمال هندوچین استقرار می‌یابد. در این زمان تاوایی منفی و مقادیر ارتفاع ژئوپتانسیل در مرکز پرفشار افزایش یافته و پرفشار

بصورت یک سلول بسته با گسترش مداری ظاهر می‌شود(شکل3b-3).

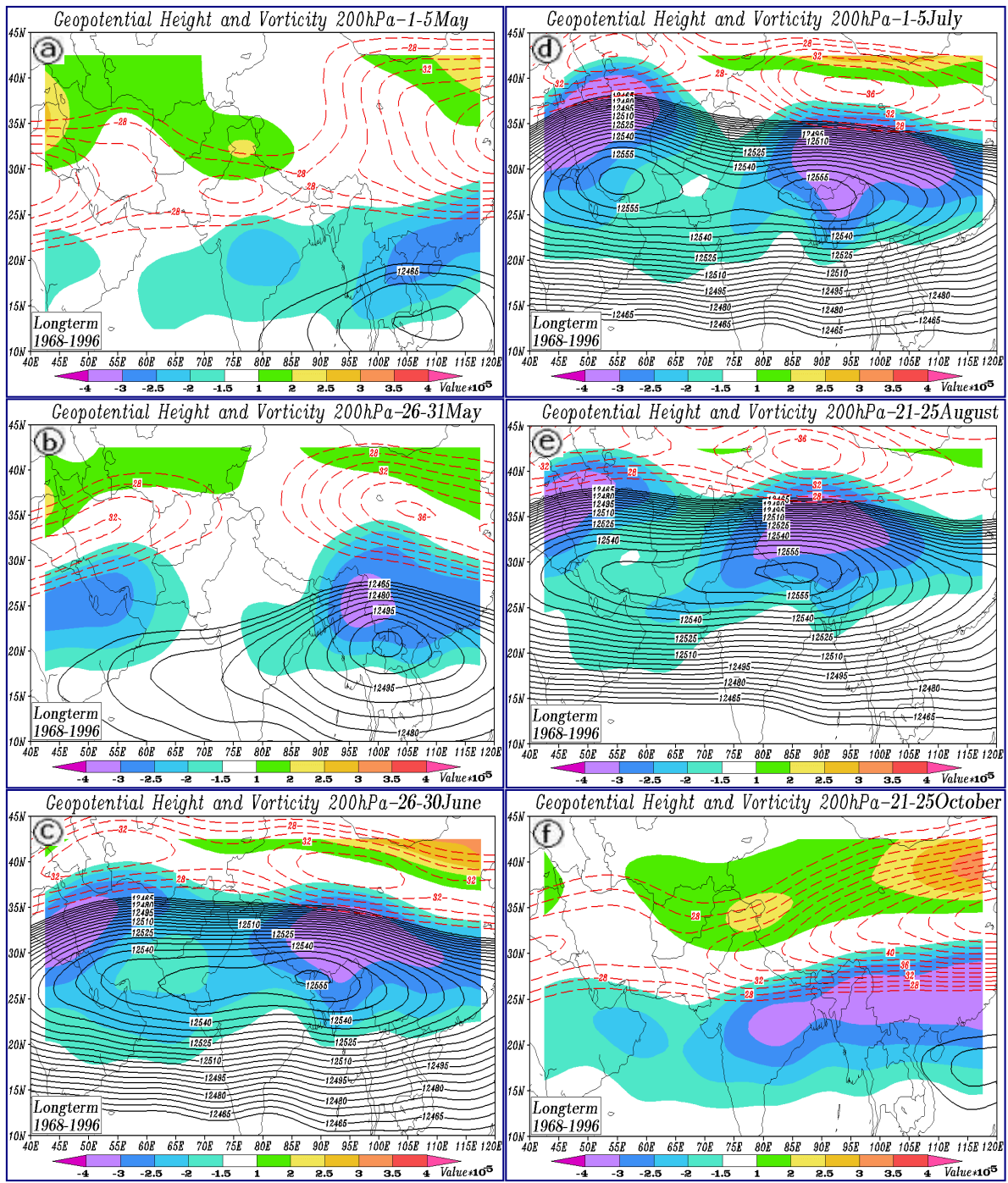

**شکل3-3.** نقشه‌های ترکیبی متوسط پنج روزه تراز 200هکتوپاسکال. تاوایی نسبی برحسب $10^{-5}\ S^{-1}$ و ارتفاع ژئوپتانسیل و سرعت باد مداری به ترتیب برحسب متر و متر بر ثانیه می‌باشد. از نشان دادن تاوایی 1/5، 1- تا 1+ ($10^{-5}\ S^{-1}$)، ارتفاع کمتر از 12460 ژئوپتانسیل متر و سرعت باد کمتر از 25 متر بر ثانیه صرفنظر شده است.

در پایان ماه ژوئن مرکز پرفشار با تداوم جابجایی شمال غرب سو، به عرض 27/5° شمالی و طول 90°

شرقی جابجا شده و در جانب جنوبی فلات مرتفع تبت استقرار می‌یابد(شکل3c-3). در این زمان بواسطه‌ی گسترش مداری پرفشار جنب حاره جنوب آسیا، منطقه‌ای از غرب اقیانوس آرام تا شمال آفریقا در جنوب عرض °35 شمالی تحت تسلط پرفشار جنب حاره است (شکل3c-3). بررسی مرکز پرفشار بیانگر آن است که در اوج تابستان در طی دوره‌ی 5 روزه نخست ماه ژوئیه و بصورت کاملاً ناگهانی مرکز پرفشاری بر روی جنوب ایران تشکیل شده و واچرخند بزرگ مستقر بر روی جنوب و جنوب غرب آسیا به شکل دو سلولی در می‌آید(شکل3d-3). ارتفاع ژئوپتانسیل در مرکز سلول پرفشار مستقر بر روی ایران به 12560 ژئوپتانسیل متر رسیده و در جانب شمال غرب و در منطقه‌ی تاج این سلول تاوایی منفی به حدود $-4\times 10^{-5}$ بر ثانیه بالغ می‌گردد. در حد فاصل دهه‌ی دوم ماه ژوئیه تا پایان دهه‌ی دوم ماه اوت گسترش مداری غرب سوی پرفشار تداوم یافته و پرفشار در منطقه‌ای حد فاصل جنوب ایران تا فلات تبت به بیشترین شدت و ارتفاع خود می‌رسد. در این زمان سلول پرفشار مستقر بر روی ایران عمدتاً کمی بالاتر از مرکز پر ارتفاع تبت ظاهر شده و در عرض °30 شمالی و طول °55 شرقی جای می‌گیرد.

در طی دوره‌ی 5 روزه‌ی 25-21 اوت، مرکز پرفشار به یکباره در طی یک جابجایی شرق سوی ناگهانی، مجدداً بر جانب جنوبی فلات تبت جای می‌گیرد(شکل3e-3). از این زمان به بعد پرفشار جنب حاره جنوب آسیا بتدریج ضعیف شده و تا پایان ماه اکتبر یک جابجایی جنوب شرق سویی را تجربه می‌کند، بطوریکه در طی دوره‌ی 5 روزه‌ی 25-21 اکتبر پرفشار بسیار ضعیف شده و در طول °12 شرقی در منتهی الیه جنوب شرق آسیا مشاهده می‌گردد (شکل3f-3). شایان ذکر است همزمان با گسترش مداری غرب سوی پرفشار در جنوب و غرب آسیا و تداوم این گسترش در طول ماه ژوئیه و اوت، دو مرکز اصلی تاوایی منفی یکی بر روی شمال غرب ایران و دیگری شمال فلات تبت شکل می‌گیرد (شکل3c-3 تا 3e-3).

**ب- پرفشار جنب حاره تابستانه در تراز میانی:** پرفشارهای جنب حاره ترازهای میانی وردسپهر مؤلفه‌ی اصلی و بلافصل کنترل کننده‌ی اقلیم تابستانه‌ی منطقه‌ی جنوب غرب آسیا به شمار می‌روند (Saaroni and Ziv, 2000; Ziv et al., 2004). بررسی متون اقلیمی بیانگر آن است که علیرغم نقش بسیار مهم پرفشارهای جنب حاره در اقلیم جنوب غرب آسیا، این مؤلفه از گردش جو کمتر مورد توجه پژوهشگران قرار گرفته و به همین جهت اطلاعات موجود در رابطه با رفتار، سازوکار تشکیل و وردش‌های مکانی و زمانی این مراکز فشار در منطقه‌ی جنوب غرب آسیا ناقص و محدود است. با توجه به اینکه الگوی استقرار و نحوه‌ی رفتار این سامانه‌های فشاری، خصوصیات اقلیمی مناطق جنب حاره جنوب غرب آسیا، از جمله وقوع یا عدم وقوع بارش‌های تابستانه در ایران را از خود متاثر می‌سازد، در ادامه با استفاده از داده های ارتفاع ژئوپتانسیل و تاوایی نسبی، وردش‌های زمانی و مکانی پرفشارهای جنب حاره تابستانه در تراز

٣٤

۵۰۰ هکتوپاسکال بر روی جنوب غرب آسیا مورد بررسی قرار می‌گیرد.

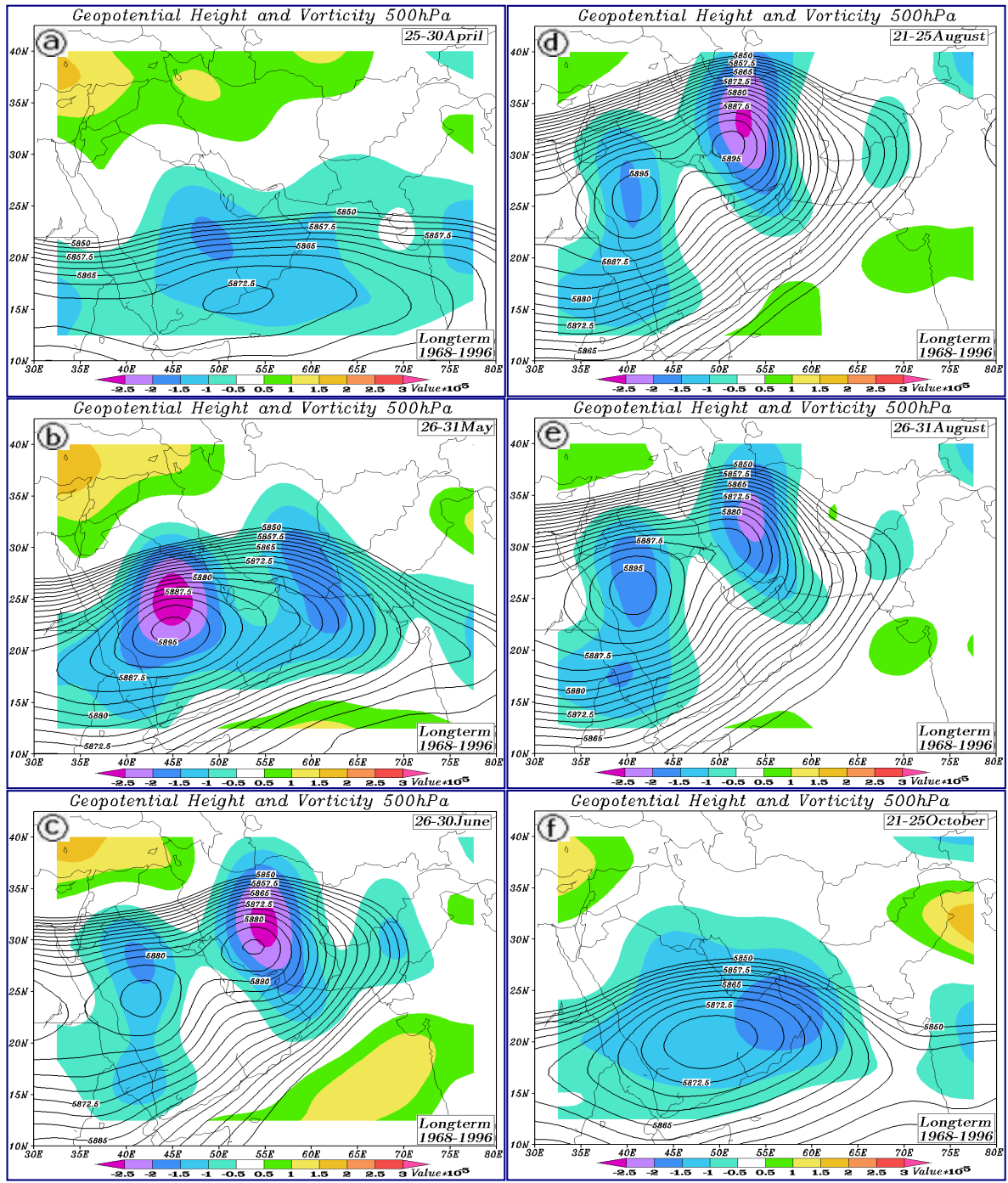

**شکل ۴-۳.** نقشه‌های ترکیبی متوسط پنج روزه تراز ۵۰۰ هکتوپاسکال. تاوایی نسبی برحسب $10^{-5}$ $s^{-1}$ و ارتفاع ژئوپتانسیل برحسب متر می‌باشد. از نشان دادن تاوایی ۰/۵- تا ۰/۵+ ($10^{-5}$ $s^{-1}$) و ارتفاع کمتر از ۵۸۵۰ ژئوپتانسیل صرفنظر شده است.

شکل ۴-۳ وردش‌های زمانی و مکانی پرفشارهای جنب حاره را در منطقه‌ی جنوب غرب آسیا نشان

همان‌طوریکه در شکل ۳-۴a مشاهده می‌شود در روزهای پایانی ماه آوریل مرکز پرفشار در منطقه‌ی جنوب غرب آسیا، بر جانب شرقی شبه جزیره‌ی عربستان در حد فاصل عرض °۱۵ شمالی و °۵۲/۵ طول شرقی استقرار می‌یابد. در این زمان مقادیر ارتفاع ژئوپتانسیل و تاوایی منفی در مرکز پرفشار پائین می‌باشد(۵۸۷۲/۵ ژئوپتانسیل متر). از این زمان تا پایان ماه می مرکز پرفشار جابجایی شمال غرب سویی را به سمت شمال شبه جزیره‌ی عربستان تجربه می‌کند، بطوریکه در طی پنج روز پایانی ماه می، مرکز پرفشار در °۴۵ طول شرقی و °۲۲/۵ عرض شمالی جای می‌گیرد. در این زمان مرکز پرفشار، در مقایسه با تمامی دوره های پنج روزه دوره گرم سال، حداکثر شدت و ارتفاع ژئوپتانسیل خود را تجربه می کند. بطوریکه مقادیر تاوایی منفی در جانب شمال سوی مرکز پرفشار به بیش از $-2/5 \times 10^{-5}$ بر ثانیه می‌رسد و ارتفاع ژئوپتانسیل در مرکز آن به ۵۸۹۵ ژئوپتانسیل متر بالغ می‌گردد (شکل ۳-۴b).

بررسی مقطع قائم جو برای دوره پنج روزه پایان ماه می، بیانگر آن است که گردش واچرخندی در شمال سوی مرکز پرفشار در تراز ۶۰۰ هکتوپاسکال به حداکثر شدت خود رسیده(شکل۳-۵b) و پرفشار در تمامی ترازهای میانی و زیرین بصورت یک سلول بسته ظاهر می شود. مرکز پرفشار فوق که از این پس آن را سلول «پرفشار عربستان» می‌نامیم تا پایان ماه ژوئن تقویت شده و کمی به سمت شمال و غرب جابجا می‌گردد، بطوریکه در پایان ماه ژوئن در عرض °۲۵ شمالی و °۴۰ طول شرقی استقرار می‌یابد(شکل۳-۴c).

نکته‌ی قابل ذکر آنکه، پرفشار عربستان در حد فاصل دوره‌ی پنج روزه‌ی پایان ماه می تا دوره‌ی پنج روزه‌ی پایان ماه ژوئن، حداکثر تاوایی منفی و بالاترین مقادیر ارتفاع ژئوپتانسیل خود را تجربه می‌کند و پس از آن در طی ماه ژوئیه و اوت از شدت گردش واچرخندی کاسته شده و ارتفاع ژئوپتانسیل درکنتور مرکزی آن کاهش می یابد. در پنج روز پایانی ماه ژوئن، سلول پرفشار دیگری در امتداد شمال شرقی پرفشار عربستان در شمال خلیج فارس بر روی ایران بسته می‌شود (شکل۳-۴c). این مرکز پرفشار که از این پس آنرا «پرفشار ایران» می‌نامیم از اواسط ماه ژوئیه تا پایان دهه‌ی دوم ماه اوت به حداکثر شدت و گستردگی خود می‌رسد. در طی این دوره، بالاترین مقادیر تاوایی منفی و ارتفاع ژئوپتانسیل در منطقه‌ی جنوب غرب آسیا بر روی ایران مشاهده می‌شود(شکل۳-۴c). پرفشار ایران که بر خلاف پرفشار عربستان مقادیر حداکثر تاوایی منفی آن بر جانب شرق-شمال شرق مرکز آن مشاهده می‌گردد، از زمان تشکیل خود در روزهای پایانی ماه ژوئن (شکل۳-۴c) تا ناپدید شدن آن در دوره‌ی پنج روزه‌ی پایان ماه اوت، کمی به سمت شمال و غرب جابجا می شود، بطوریکه مرکز پرفشار در اغلب روزهای تابستان بطور متوسط در طول °۵۰ شرقی و عرض °۳۲/۵ شمالی و حداکثر تاوایی منفی آن در طول °۵۲/۵ شرقی (شکل۳-۵a) مشاهده می گردد. بررسی تطبیقی مقادیر ارتفاع ژئوپتانسیل و تاوایی نسبی در سلول پرفشار ایران،

نشان‌دهنده آن است که مقادیر تاوایی $-2/5 \times 10^{-5}$ بر ثانیه می‌تواند معیار مناسبی جهت تشخیص تکوین زمانی و مکانی پرفشار ایران در تراز ۵۰۰ هکتوپاسکال باشد(شکل۵-۳a ).

در دوره‌ی پنج روزه‌ی پایانی ماه اوت مرکز پرفشار ایران علیرغم گردش واچرخندی قوی و ارتفاع ژئوپتانسیل بسیار بالا(تاوایی $-2/5 \times 10^{-5}$ و ارتفاع ۵۸۹۵ ژئوپتانسیل متر) به یکباره از شمال خلیج فارس محو می‌گردد(شکل۴-۳e). این در حالیست که مرکز پرفشار عربستان در همین دوره مجدداً تقویت شده و ارتفاع ژئوپتانسیل و مقادیر تاوایی منفی در مرکز آن افزایش می‌یابد(شکل۴-۳e). بدین ترتیب الگوی دو سلولی پرفشارجنب حاره در جنوب غرب آسیا تنها به مدت دو ماه(۲۵-۳۰ ژوئن تا ۳۱-۲۵ اوت) تداوم یافته و پس از آن به الگوی تک سلولی پرفشار عربستان در جنوب مبدل می‌گردد. البته لازم به یادآوری است که گردش واچرخندی بر روی نیمه غربی ایران تا اواسط ماه سپتامبر تداوم می‌یابد(شکل۵-۳a). مرکز پرفشار عربستان تا پایان سپتامبر در عرض ۲۵° شمالی باقی‌مانده اما پس از آن تا پایان اکتبر طی یک جابجایی جنوب سوی شرق تدریجی و کند، به موقعیت آغازین خود در فصل بهار باز می‌گردد. بطوریکه با توجه به شکل۴-۳f در طی دوره‌ی پنج روزه‌ی پایانی ماه اکتبر در ۵۰° طول شرقی و ۲۰° عرض شمالی جای می‌گیرد. در این زمان مقادیر گردش واچرخندی و ارتفاع ژئوپتانسیل در مرکز پرفشار عربستان در مقایسه با فصل تابستان کاهش یافته است (شکل۴-۳f).

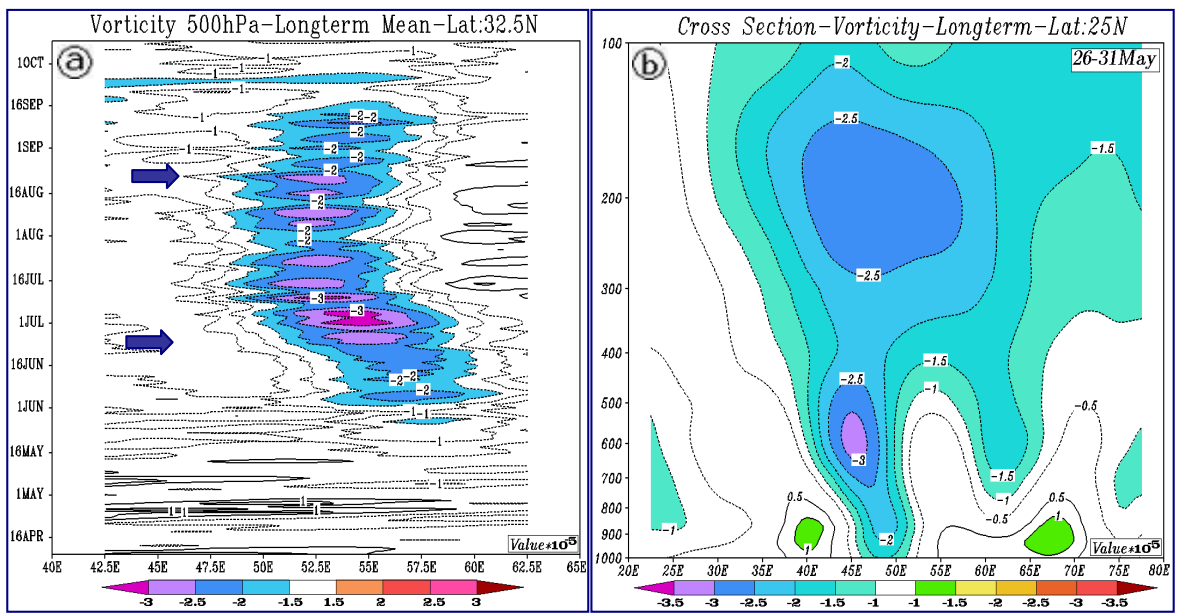

**شکل۵-۳.** (a) مقادیر تاوایی نسبی از ۱۵ آوریل تا ۱۵ اکتبر در عرض ۳۲/۵° شمالی. مقادیر تاوایی $-2/5 \times 10^{-5}$ بر ثانیه با دوره تشکیل و استقرار مرکز پرفشار بر روی ایران انطباق دارد. (b) مقطع قائم تاوایی در امتداد عرض ۲۷/۵° شمالی برای زمان اوج شدت پرفشار عربستان(دوره پنج روزه ۳۱-۲۶ می) برحسب $10^{-5}\ s^{-1}$.

در بررسی تطبیقی موقعیت و شدت مراکز پرفشار ترازهای میانی و فوقانی بر روی جنوب و جنوب غرب آسیا نکته‌ی جالبی آشکار می‌گردد. بدین ترتیب که روند جابجایی پرفشار عربستان از ماه آوریل تا پایان ماه ژوئن در تراز میانی با روند جابجایی شمال سوی مرکز پرفشار تراز فوقانی در جنوب آسیا انطباق خوبی را نشان می‌دهد(شکل3-3a تا 3-3c و شکل3-4a تا 3-4c). در مقابل عقب نشینی جنوب شرق سوی مراکز پرفشار فوقانی و میانی از روزهای پایانی ماه اوت تا پایان اکتبر نیز از چنین انطباقی برخوردار است(شکل3-3d تا 3-3e و شکل3-4d تا 3-4e). نکته‌ی جالب‌تر آنکه شکل‌گیری ناگهانی پرفشار ایران در تراز میانی در ابتدای ماه ژوئیه با گسترش غرب سو و شکل‌گیری سلول پرفشار تراز فوقانی بر روی ایران در امتداد عرض °30 شمالی و در عین حال ناپدید شدن پرفشار تراز فوقانی با ناپدید شدن سلول پرفشار ایران در تراز میانی انطباق بسیار خوبی را نشان می‌دهد(شکل3-3c تا 3-3e و شکل3-4c تا 3-4e).

با توجه به نحوه‌ی جابجایی و میزان شدت پرفشارها، بنظر می‌رسد تشکیل و جابجایی مراکز پرفشار بر روی شبه جزیره عربستان و جنوب آسیا به ترتیب در ترازهای میانی و فوقانی وردسپهر از آهنگ فصلی مشابهی تبعیت می‌نمایند. در حالیکه شکل‌گیری و تکوین پرفشارها در ترازهای فوق بر روی ایران از الگوی متمایزی برخوردار است. چرا که شکل‌گیری و ناپدید شدن ناگهانی مراکز پرفشار بر روی ایران علیرغم تداوم شدت گردش واچرخندی و ارتفاع ژئوپتانسیل بالا در مرکز این پرفشارها در تمام مدت تشکیل، با آهنگ فصلی تابش و گردش جو مرتبط با آن انطباق ندارد.

### 3-2-1-2. جت جنب حاره بر روی جنوب غرب آسیا

در طی چند دهه‌ی گذشته پژوهشگران متعددی سعی نموده‌اند تا نقش جت جنب حاره را بر روی اقلیم منطقه‌ی جنوب و جنوب شرق آسیا شناسایی نمایند. در یکی از نخستین پژوهش‌ها «ین[34]» در سال 1949، در رابطه با موضوع جالب و بحث‌انگیزی چون «آغازگری» موسمی تابستانه بر روی هند، تئوریی را مطرح نمود مبنی بر اینکه جابجایی ناگهانی و شمال سوی جت جنب حاره از جنوب رشته کوه هیمالیا به جانب شمالی آن در ابتدای ماه ژوئن، استقرار گردش تابستانه و آغازگری موسمی هند را بدنبال دارد. بدنبال آن یافته‌های پژوهشگران چینی در رابطه با تعیین و تقسیم فصول طبیعی در جنوب و بویژه شرق آسیا بر اساس وردش فصلی جت جنب حاره‌ی غربی(Staff Members, 1957; 1958; Dao and Chen, 1957)، انگیزه‌ی مطالعه‌ی جت جنب حاره را در وردسپهر فوقانی بر روی مناطق مختلف آسیا افزایش داد ( Krishnamurti,

---

34.Yin

1961; Yang and Webster, 1990; Kuang and Zhang, 2005). بدنبال تحقیقات فوق، مطالعات متعدد و بی‌شماری نقش جت جنب حاره را بر خصوصیات اقلیمی مناطق جنوب-جنوب شرقی آسیا در فصل موسمی تابستانه مورد توجه قرار دادند و ارتباط وردش‌های زمانی و مکانی جت با آغازگری و شدت موسمی در جنوب آسیا (Webster and Yang, 1992; Yang et al., 2004)، توزیع مکانی و زمانی بارش در جنوب و شرق آسیا (Liang and Wang,1998; Yang et al.,2002; Lu,2004) و همچنین نقش جت در تغییر فصل (Sutcliffe and Bannon, 1956; Yeh, 1959) را تبیین نمودند.

مطالعات اولیه در رابطه با گردش بزرگ مقیاس جو در منطقه‌ی آسیا بیانگر آن است که تغییر ناگهانی گردش فصلی در ماه ژوئن در جنوب و غرب آسیا با جابجایی ناگهانی شمال سوی جت غربی بر فراز خاورمیانه در ابتدای ماه ژوئن مرتبط است ( Dao and Chen, 1957; Sutcliffe and Bannon, 1956; Yeh, 1959). همچنین بررسی‌های اخیر جهت پیش‌بینی موسمی تابستانه در جنوب آسیا، جت جنب حاره‌ی خاورمیانه را بعنوان یک پیش‌بین مناسب معرفی نموده‌اند(Webster and Yang, 1992; Yang et al., 2004). علیرغم مطالعات انجام شده، ساختار و خصوصیات جت جنب حاره‌ی تابستانه بر روی جنوب غرب آسیا کمتر مورد توجه قرار گرفته است. در ادامه خصوصیات و ویژگی‌های جت جنب حاره تابستانه بر روی جنوب غرب آسیا مورد بررسی قرار می‌گیرد. برای این منظور از مؤلفه‌ی باد مداری برای یک دوره‌ی اقلیمی ۲۹ ساله(۱۹۶۸-۱۹۹۶) استفاده شده است.

شکل های ۳-۶a تا ۳-۶f سرعت متوسط باد را در مقطع قائم بر روی جنوب غرب آسیا در امتداد نصف النهار ۵۵° شرقی نشان می‌دهند. با توجه به شکل‌ها، بالاترین میزان سرعت جت جنب حاره در تراز ۲۰۰ هکتوپاسکال مشاهده می‌شود. بررسی زمانی موقعیت و شدت جت جنب حاره نشاندهنده‌ی آن است که در دوره‌ی پنج روزه‌ی اول ماهِ آوریل، هسته‌ی جت در حول و حوش عرض ۲۵° شمالی جای می‌گیرد(شکل ۳-۶a). جت غربی تا پایان دهه‌ی اول ماه می جابجایی چندانی از خود نشان نمی‌دهد و در عرض ۲۵° تا ۲۷° شمالی استقرار می‌یابد. اما در حد فاصل دهه‌ی دوم ماه می(شکل۳-۶b) تا دهه اول ماه ژوئن، جت بر روی جنوب غرب آسیا به یکباره حدود ۸ درجه به سمت شمال جابجا شده و در موقعیت تابستانی خود قرار می‌گیرد(شکل۳-۶d). مقایسه‌ی مقطع قائم سرعت باد مداری برای دوره‌ی فوق، بخوبی این نکته را نشان می‌دهد(شکل۳-۶b تا ۳-۶d). در ماه ژوئیه با شکل‌گیری مرکز پرفشار جنب حاره بر روی ایران، جت جنب حاره جنوب غرب آسیا به شمالی‌ترین موقعیت خود در طول سال منتقل می‌گردد (شکل۳-۶e و ۳-۷). در این زمان در منطقه‌ی آسیا دو هسته سرعت در جت جنب حاره، یکی بر روی دریای خزر و دیگری در جانب قطب سوی فلات تبت مشاهده می‌گردد(شکل۳-۳c و ۳-۳e). شایان ذکر

است که حداکثر متوسط سرعت در هسته‌ی جت تابستانه بر روی جنوب غرب آسیا بین ۳۲ تا ۳۶ متر بر ثانیه می‌باشد.

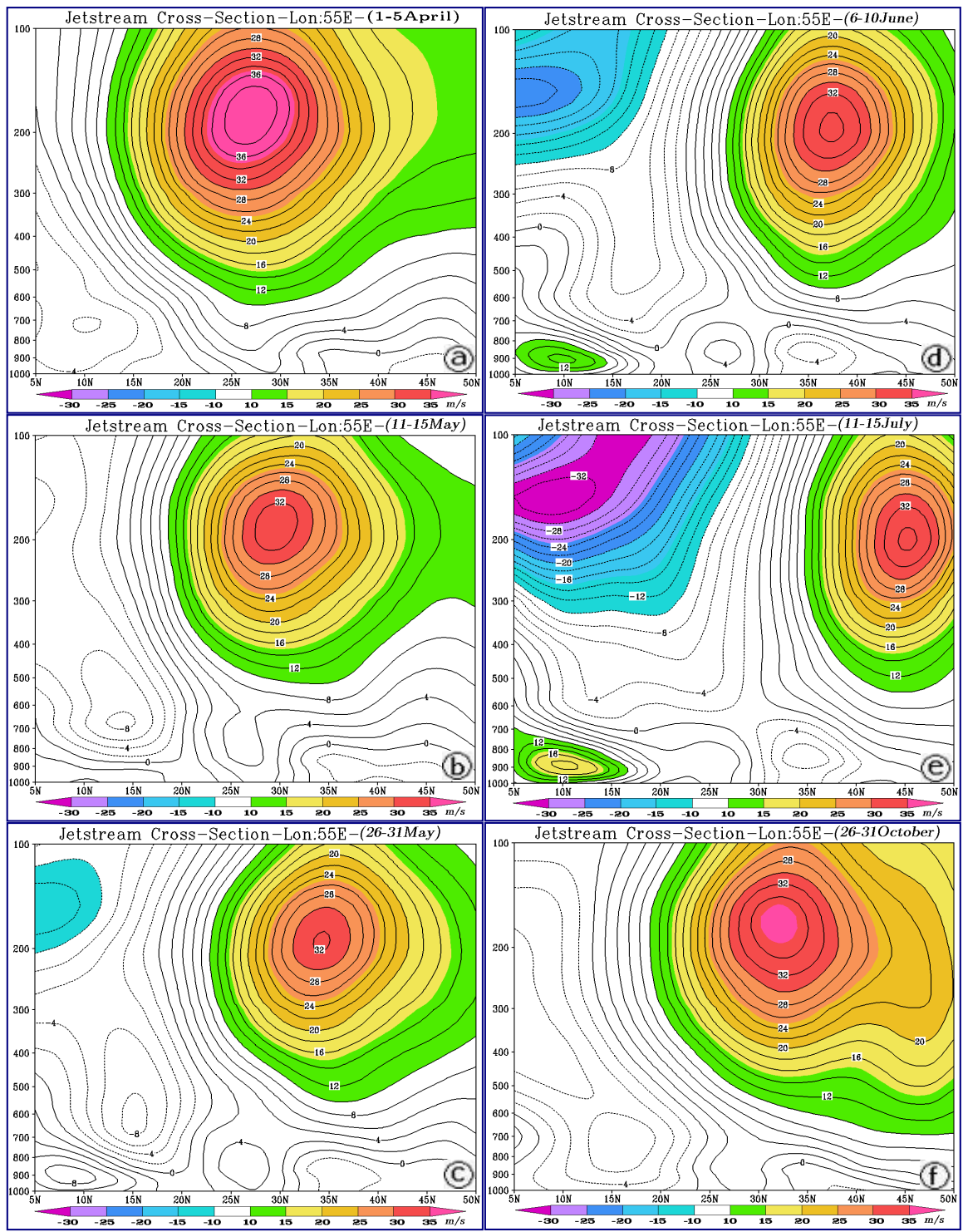

**شکل ۳-۶.** مقطع قائم باد مداری در امتداد نصف النهار ۵۵° شرقی برای دوره های زمانی پنج روزه برحسب متر بر ثانیه. باد غربی با خط ممتد و باد شرقی با خط چین به نمایش درآمده است.

بررسی متوسط پنج روزه‌ی جت جنب حاره نکات بسیار جالبی را آشکار می‌سازد. بدین ترتیب که در حد فاصل پنج روز پایانی ماه می تا ابتدای ماه ژوئن، علاوه بر تقویت جت جنب حاره و جابجایی شمال سوی ناگهانی آن در منطقه‌ی جنوب غرب آسیا، جت شرقی حاره‌ای در تراز فوقانی و جت غربی موسوم به جت سومالی در تراز زیرین به یکباره بر روی غرب اقیانوس هند پدیدار شده و به شدت تقویت می‌گردند. با توجه به شکل‌3-6c و 3-6d جت شرقی در ابتدای ماه ژوئن، بطور ناگهانی گسترش یافته و سرعت در هسته‌ی آن در تراز 150 هکتوپاسکال به میزان 22 متر بر ثانیه بالغ گردیده و سرعت درجت سومالی نیز به 16 متر بر ثانیه می‌رسد. بدین ترتیب با توجه به الگوی متوسط جریان هوا، جابجایی ناگهانی شمال سوی جت جنب حاره همراه با پدیدار شدن و تقویت ناگهانی جت شرقی در تراز فوقانی و جت سومالی در تراز زیرین در دوره‌ی پنج روزه‌ی دوم ماه ژوئن، نشان‌دهنده‌ی استقرار رژیم گردش بزرگ مقیاس تابستانه در این زمان از سال بر روی جنوب غرب آسیاست.

نتایج این بررسی با یافته‌های پیشین در رابطه با آغاز فصل تابستان بر روی جنوب و شرق آسیا ( Yin, 1949; Staff Members, 1957; 1958; Dao and Chen, 1957; Yeh, 1959 ) انطباق خوبی را نشان می‌دهد. جهت بررسی نحوه تکوین جت جنب حاره بر روی جنوب غرب آسیا، «نمودار هاومولر»[35] برای تراز 200 هکتوپاسکال در امتداد نصف النهار °55 شرقی تهیه شد(شکل3-7). بررسی روند جابجایی و میزان سرعت در هسته‌ی جت جنب حاره در نمودار هاومولر نشان‌دهنده‌ی آن است که از اواسط ماه می تا نیمه‌ی ماه ژوئیه، جت جنب حاره یک جابجایی شمال سوی ممتد و در عین حال پرشیبی را بر روی ایران طی می‌کند. بدین ترتیب موقعیت متوسط محور جت از حول و حوش عرض °30 در 15می به عرض °45 شمالی در پایان دهه اول ژوئیه انتقال می‌یابد(شکل3-7).

با توجه به شکل، جت جنب حاره به موازات شکل گیری و گسترش مرکز پرفشار جنب حاره بر روی ایران، در شمالی ترین موقعیت خود استقرار می یابد. بطوریکه محور جت جنب حاره از نیمه‌ی ماه ژوئیه تا پایان ماه اوت در عرض °45 شمالی باقی مانده و از حداکثر شدت خود برخوردار است. در این زمان جت شرقی نیز حداکثر گستردگی و شدت خود را تجربه می کند. اما با آغاز ماه سپتامبر و بدنبال ناپدید شدن پرفشار جنب حاره ایران، جت جنب حاره روند جابجایی استوا سوی آرامی را آغاز نموده و تا پایان ماه اکتبر در حول و حوش عرض °32/5 شمالی جای می گیرد(شکل3-7).

---

35. Hovmoller Diagram

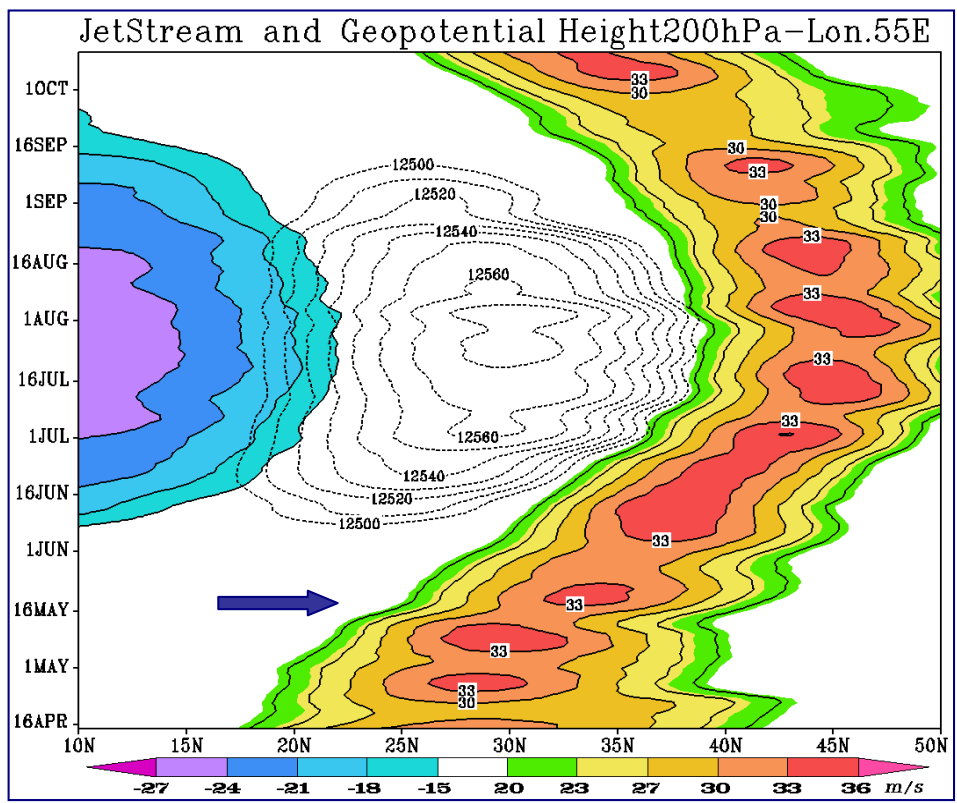

**شکل۷-۳.** موقعیت و شدت جت غربی، جت شرقی و پرفشار جنب حاره از ۱۵ آوریل تا ۱۵ اکتبر در امتداد طول ۵۵° شرقی. از نشان دادن سرعت باد مداری ۱۵- تا ۲۰متر بر ثانیه و ارتفاع کمتر از ۱۲۵۰۰ ژئوپتانسیل متر صرفنظر شده است. پیکان ضخیم زمان جابجایی ناگهانی جت جنب حاره را نشان می‌دهد.

## ۲-۲-۳. گردش جو تابستانه در ترازهای زیرین جو

بررسی متون اقلیمی بیانگر آن است که ویژگی‌های گردش جو در ترازهای زیرین و بر روی جنوب غرب آسیا کمتر مورد توجه پژوهشگران قرار گرفته و در این میان بررسی ساختار گردش جو در فصل تابستان از کمترین اقبال برخوردار بوده است(Arakawa and Takahashi, 1981). بر این اساس، به منظور شناسایی مؤلفه‌های اصلی گردش جو تابستانه در ترازهای زیرین بر روی جنوب غرب آسیا وضعیت ارتفاع ژئوپتانسیل، تاوایی و مقادیر اُمگا برای ترازهای ۸۵۰ و ۷۰۰ هکتوپاسکال در دوره‌های زمانی مختلف مورد بررسی قرار گرفت. بررسی نقشه‌های متوسط ۲۹ ساله (۱۹۹۶-۱۹۶۸) ارتفاع ژئوپتانسیل در تراز ۸۵۰ هکتوپاسکال برای ماه‌های ژوئن، ژوئیه و اوت، تسلط یک مرکز کم‌فشار بر روی پاکستان و یک زبانه‌ی کم‌فشار با کشیدگی غرب سو بر روی نیمه‌ی غربی ایران را نشان می‌دهد(شکل ۳-۸a تا ۳-۸c). با توجه به شکل‌ها این دو مرکز کم‌فشار، تاوایی مثبت و گردش چرخندی گسترده‌ای را در امتداد مداری ایجاد می‌نمایند. در مقابل، وجود دو مرکز گردش واچرخندی در شرق دریای خزر و غرب خلیج‌فارس الگوی استقرار این مراکز کم‌فشار را در ماه‌های تابستان از خود متأثر ساخته و الگوی گردش پیچیده‌ای را بر روی

۴۲

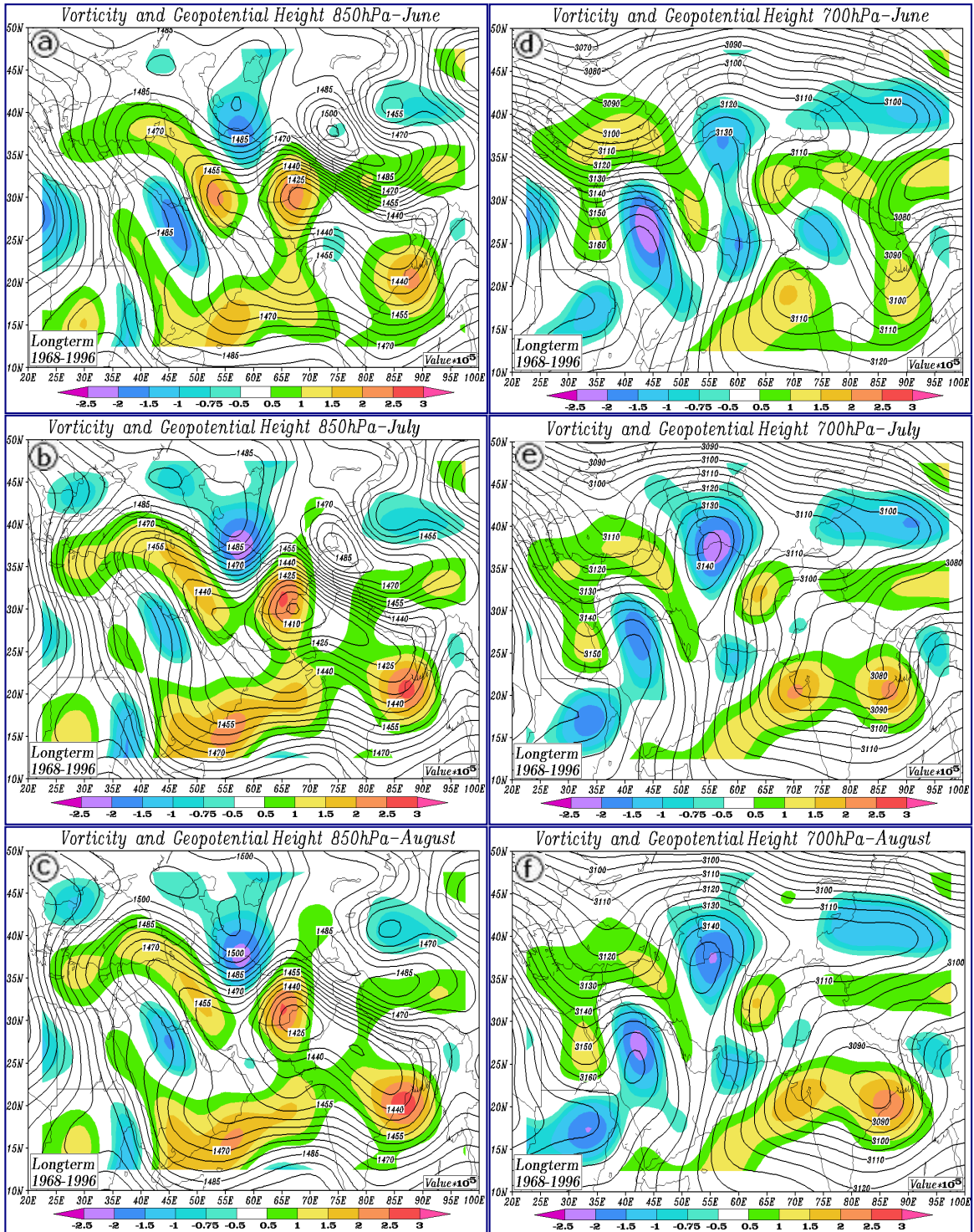

**شکل 3-8.** وضعیت متوسط ماهانه ارتفاع و تاوایی نسبی در ماه‌های ژوئن، ژوئیه و اوت در جنوب و جنوب غرب آسیا. (a، b و c) تراز 850 هکتوپاسکال. (d، e و f) تراز 700 هکتوپاسکال. تاوایی برحسب $10^{-5}\,s^{-1}$ و ارتفاع ژئوپتانسیل برحسب متر بوده و از نشان دادن تاوایی 0/5- تا 0/5+ ($10^{-5}\,s^{-1}$) صرف‌نظر شده است.

بررسی نقشه‌های تراز ۷۰۰ هکتوپاسکال و مقایسه‌ی آن با تراز ۸۵۰ هکتوپاسکال نشان‌دهنده‌ی آن است که با افزایش ارتفاع، بر شدت گردش واچرخندی وگستردگی مراکز پرفشار شرق دریای خزر و غرب خلیج‌فارس(مرکز پرفشار عربستان) افزوده می‌شود. بطوریکه بر روی نقشه‌ها، مرکز پرفشار شرق دریای خزر بخصوص در ماه‌های ژوئیه و اوت با مقادیر تاوایی منفی بیش از $-2\times10^{-5}$ بر ثانیه در مرکز و کنتور بسته‌ی مرکزی ۳۱۴۰ ژئوپتانسیل متر کاملاً برجسته و مشخص است(شکل۳-۸e تا ۳-۸f). در مقابل، مرکز گردش واچرخندی غرب خلیج‌فارس(پرفشار عربستان) نیز در مقایسه با تراز ۸۵۰ هکتوپاسکال تقویت شده و در طول ماه‌های ژوئن و ژوئیه بصورت یک پشته(شکل۳-۸d و۳-۸e) و در ماه اوت به شکل یک مرکز پر ارتفاع با کنتور بسته‌ی مرکزی ۳۱۶۵ ژئوپتانسیل متر خودنمائی می‌کند (شکل۳-۸f). در این تراز مرکز کم‌فشار پاکستان به شدت ضعیف شده، بطوریکه بدون هیچ کنتور ارتفاعی بسته‌ای تنها به صورت یک منطقه تاوایی مثبت ضعیف قابل مشاهده است. زبانه‌ی کم فشار غرب ایران نیز ناپدید شده و همچون کم فشار پاکستان تنها با تاوایی مثبت ضعیفی در امتداد رشته کوه زاگرس در حدفاصل شمال غرب ایران تا خلیج فارس قابل شناسایی است.

نکته‌ی برجسته‌ی نقشه‌ی تراز ۷۰۰ هکتوپاسکال، استقرار یک ناوه‌ی نسبتاً عمیق در تمامی ماه‌های تابستانی در شرق ترکیه است. بنظر می‌رسد عمیق شدن ناوه‌ی تراز ۷۰۰ هکتوپاسکال در حد فاصل شرق ترکیه تا شمال غرب ایران، زبانه‌ی کم‌فشار تراز ۸۵۰ هکتوپاسکال را در نیمه غربی ایران تقویت نموده و کشیدگی و گسترش غرب سوی آن را موجب گردیده است. بدین ترتیب مؤلفه‌های اصلی گردش جو در ترازهای زیرین جو بر روی جنوب غرب آسیا عبارتند از:

- واچرخند ترکمنستان
- ناوه‌ی شبه ساکن شرق ترکیه
- کم‌فشار پاکستان
- زبانه‌ی کم‌فشار زاگرس
- پرفشار عربستان.

بررسی سازوکار وقوع بارش‌های تابستانه‌ی فلات ایران، نقش بسیار مهم مراکز فشار ترازهای زیرین جو را در وقوع این بارش‌ها نشان می‌دهد.

به همین جهت در ادامه، مؤلفه‌های اصلی گردش جو در ترازهای زیرین بر روی جنوب غربی آسیا بطور جداگانه مورد تجزیه و تحلیل قرار می‌گیرد.

٤٤

## 1-2-2-3- واچرخند ترکمنستان

در سال 1996 «هاسکینز»[36] با انتشار مقاله‌ای تحت عنوان «وجود و تشدید واچرخند های جنب حاره تابستانه» در بولتن انجمن هواشناسی آمریکا و همچنین با چاپ مقاله‌ای به همراه «رادول»[37] تحت عنوان «موسمی‌ها و دینامیک بیابان‌ها»، چالش بزرگی را در حوزه‌ی اقلیم‌شناسی در رابطه با واچرخندهای جنب حاره تابستانه ایجاد نمود. سؤال مهمی که هاسکینز درصدد پاسخ‌گویی به آن برآمد، این بود که «چرا واچرخندهای جنب حاره تابستانه در نیمکره شمالی از قرینه زمستانی خود قوی‌ترند؟» در طی چند سال گذشته پژوهشگران بی‌شماری درصدد پاسخ‌گویی به این پرسش برآمده‌اند (Rodwell and Hoskins, 1996; 2001; Hoskins et al., 1999; Chen et al., 2001; Zhang et al., 2002; Wu and Liu, 2003; Liu and Wu, 2004; Liu et al., 2004; Ziv et al., 2004; Chen, 2005; Nakamura and Miyasaka, 2005) و سازوکارهای متعددی جهت تشکیل و تقویت واچرخندهای جنب حاره تابستانه ارائه نموده اند (Wu et al., 2004). از جمله نتایج این پژوهش‌ها، کشف یک منطقه‌ی نزول هوا در حدفاصل دریاچه‌ی آرال تا شرق دریای خزر است. توضیح بیشتر اینکه، رادول و هاسکینز در مقالات متعددی ضمن کشف دو منطقه‌ی نزول محلی یکی بر شرق دریای خزر و دیگری در مدیترانه‌ی شرقی، تشکیل واچرخندهای جنب حاره تابستانه در مناطق فوق را ناشی از هوای با دمای پتانسیل بالا با منشأ موسمی هند دانسته‌اند (Rodwell and Hoskins, 1996; 2001; Hoskins et al., 1999). نکته‌ی جالب اینکه با توجه به بررسی‌های انجام شده توسط نگارنده، در هیچ یک از متون اقلیمی مربوط به منطقه‌ی جنوب غرب آسیا، اشاره‌ای به وجود چنین مرکز پرفشاری نشده است. در ادامه، خصوصیات واچرخند ترکمنستان مورد بررسی قرار می‌گیرد.

جهت شناسایی واچرخند ترکمنستان ابتدا نقشه‌های ماهانه‌ی ترازهای مختلف جو مورد بررسی قرار گرفت. نتایج بررسی نشان داد که واچرخند ترکمنستان در تراز 700 هکتوپاسکال ابتدا در طول ماه ژوئن بصورت یک پشته بر جانب شرقی خزر با حداکثر تاوایی منفی 0/00002- بر ثانیه ظاهر شده (شکل 3-8d) و سپس در طی ماه ژوئیه و اوت بصورت یک واچرخند بسته در عرض °37/5 شمالی و در حدفاصل °57 تا °60 طول شرقی استقرار می‌یابد. در این زمان میزان گردش واچرخندی در مرکز پرفشار افزایش یافته و مقدار تاوایی به 0/000025- بر ثانیه می‌رسد (شکل 3-8e تا 3-8f). در نقشه‌های متوسط ماهانه‌ی مربوط به ماه‌های ژوئیه و اوت مرکز واچرخند عمدتاً با کنتور بسته‌ی 3140 ژئوپتانسیل متر در جانب شرقی دریای خزر مشخص می‌گردد (شکل 3-8e).

---

36. Hoskins

37. Rodwell

بررسی میزان سرعت قائم (اُمگا) در مرکز واچرخند در مقطع قائم جو، برای ماه‌های ژوئن، ژوئیه و اوت نشان‌دهنده آن است که نزول هوا از بالای تراز ۲۰۰ هکتوپاسکال آغاز شده و تا تراز ۹۰۰ هکتوپاسکال ادامه می‌یابد(شکل d۹-۳).

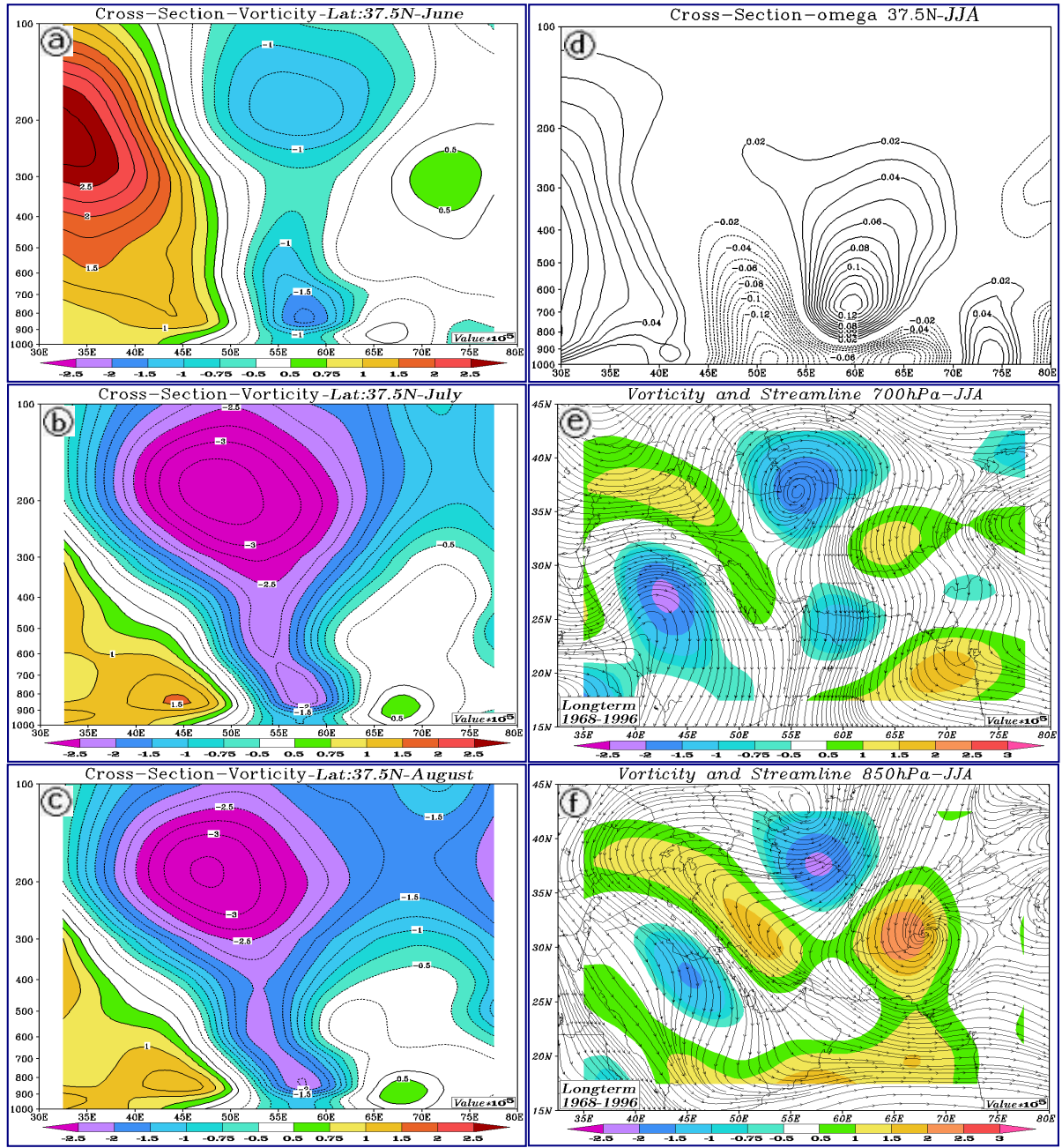

**شکل ۹-۳.** (a، b و c) متوسط ماهانه تاوایی نسبی در مقطع قائم جو در امتداد عرض ۳۷/۵° شمالی به ترتیب برای ماه‌های ژوئن، ژوئیه و اوت. (d) متوسط فصلی اُمگا در امتداد عرض ۳۷/۵° شمالی در مقطع قائم جو. (e و f) تاوایی نسبی و الگوی گردش متوسط فصلی به ترتیب در تراز ۷۰۰ (e) و ۸۵۰ هکتوپاسکال(f). اُمگا بر حسب Pa/s و تاوایی برحسب $s^{-1}$ $10^{-5}$ است. کلیه داده‌ها مربوط به یک دوره ۲۹ ساله(۱۹۹۶-۱۹۶۸) می‌باشد.

با توجه به شکل 3-9d حداکثر نزول هوا در واچرخند ترکمنستان در تراز 700 هکتوپاسکال بوقوع می‌پیوندند. شدت نزول هوا در این تراز در مقیاس زمانی درازمدت (29ساله) به 0/13 پاسکال بر ثانیه بالغ می‌گردد. جهت تعیین شدت گردش درواچرخند ترکمنستان مقادیر متوسط ماهانه تاوایی نسبی در مقطع قائم جو مورد بررسی قرار گرفت. بررسی مقادیر متوسط ماهانه‌ی تاوایی در مقطع قائم جو، بیانگر آن است که واچرخند ترکمنستان در ماه ژوئن در محدوده‌ی °58 طول شرقی در تراز 800 هکتوپاسکال، حداکثر گردش واچرخندی خود را تجربه می‌کند (شکل 3-9a). در این ماه همانطوریکه شکل 3-9a نشان می‌دهد با وجود شکل‌گیری گردش واچرخندی (تاوایی منفی) در تمام ستون جو، حداکثر مقدار تاوایی منفی به میزان $-1/5 \times 10^{-5}$ تا $-1/75 \times 10^{-5}$ بر ثانیه در حدفاصل تراز 850 تا 700 هکتوپاسکال مشاهده می‌گردد. در طی ماه‌های ژوئیه و اوت، بواسطه‌ی استقرار یک واچرخند بزرگ بر روی منطقه‌ی جنوب غرب آسیا، تاوایی منفی در ترازهای فوقانی وردسپهر بطور قابل ملاحظه‌ای افزایش یافته، بطوریکه مقادیر تاوایی منفی در تراز 200 هکتوپاسکال در عرض °37/5 شمالی و طول °55 شرقی به $-3/25 \times 10^{-5}$ بر ثانیه بالغ می‌گردد(شکل 3-9b و 3-9c). در این زمان نزول دینامیکی هوا در ستونی نسبتاً باریک در حدفاصل °55 تا °60 طول شرقی و عرض °37/5 شمالی، موجب تقویت واچرخند ترکمنستان شده و میزان گردش واچرخند را در مرکز پرفشار افزایش می‌دهد(شکل 3-9b و 3-9c).

با توجه به شکل 3-9b، تاوایی منفی در ماه ژوئیه در حدفاصل تراز 800 تا 600 هکتوپاسکال به حداکثر خود یعنی $-2/25 \times 10^{-5}$ بر ثانیه می‌رسد. در ماه اوت این میزان تاوایی تنها در تراز 800 هکتوپاسکال قابل مشاهده است(شکل 3-9c).

جهت بررسی تکوین زمانی واچرخند ترکمنستان، نمودار هاومولر از مقادیر اُمگا برای تراز 700 هکتوپاسکال تهیه گردید. با در نظر گرفتن شکل 3-10، نزول هوا در منطقه‌ی اصلی استقرار واچرخند ترکمنستان از اواسط ماه می آغاز شده و تا ابتدای ماه اکتبر ادامه می‌یابد. در این میان شدت نزول هوا در حدفاصل دوره پنج روزه اول ماه ژوئن تا روز 16 سپتامبر به حداکثر میزان خود می‌رسد(این محدوده‌ی زمانی در شکل 3-10 با اُمگای 0/10 پاسکال برثانیه و با رنگ قرمز مشخص گردیده است). با در نظر گرفتن مقادیر متوسط روزانه اُمگا در مرکز واچرخند، پرفشار ترکمنستان در حدفاصل زمانی روز اول ژوئیه تا 20 اوت، شدیدترین نزول هوا را در مرکز خود تجربه می‌کند. بطوریکه مقادیر متوسط اُمگا از 0/125 پاسکال برثانیه تجاوز می‌کند(شکل 3-10). بررسی‌های انجام شده نشان می‌دهدکه واچرخند ترکمنستان بعنوان یک مؤلفه‌ی مقیاس سینوپتیک با توجه به موارد زیر گردش جو بر روی منطقه‌ی جنوب غرب آسیا را از خود متأثر می‌سازد:

۱- بررسی‌های انجام شده نشانگر آن است که استقرار واچرخند ترکمنستان، شکل‌گیری و تداوم ناوه‌ی شبه‌ایستای شرق ترکیه را بدنبال دارد. بدین ترتیب که واچرخند ترکمنستان پس از شکل‌گیری در جانب شرقی دریای خزر بصورت یک مانع جوی در برابر جریانات شرق سو عمل نموده و شکل‌گیری ناوه‌ای در غرب خود موجب می‌گردد. بررسی نقشه‌های پنج روزه‌ی تراز ۷۰۰ هکتوپاسکال از ابتدای تشکیل واچرخند ترکمنستان تا زمان ناپدید شدن آن بخوبی بیانگر این موضوع است.

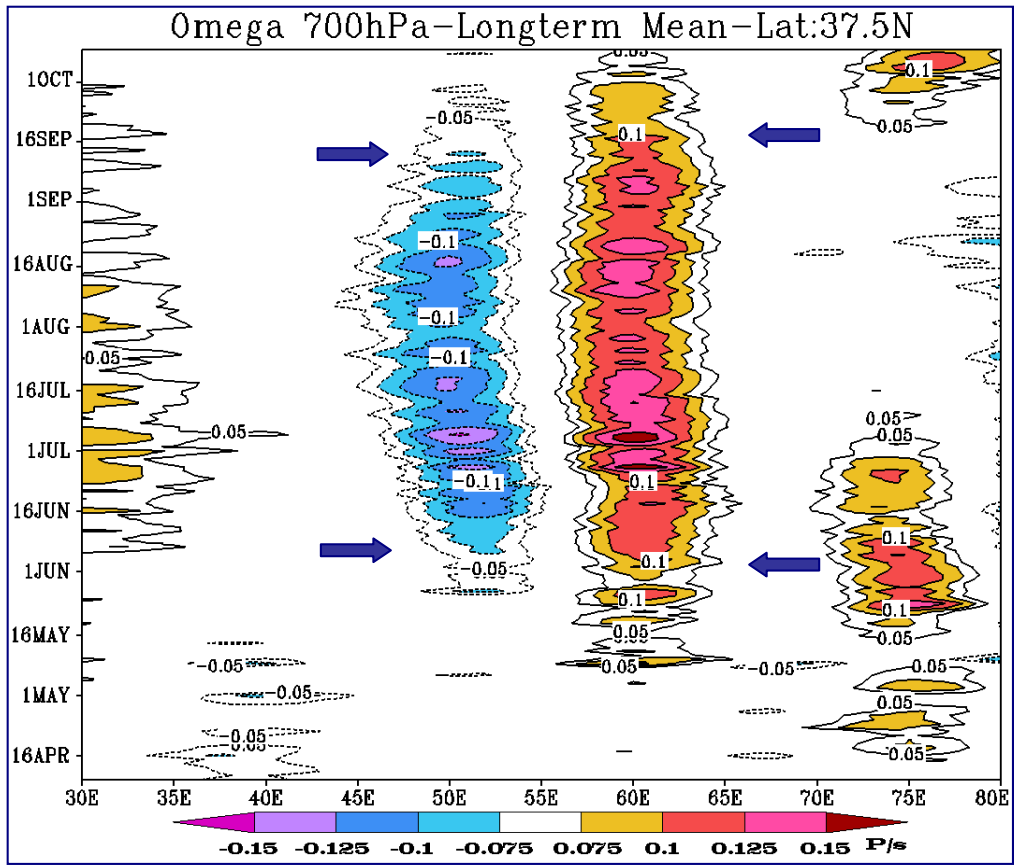

**شکل ۳-۱۰.** مقادیر متوسط روزانه درازمدت(۱۹۶۸-۱۹۹۶) اُمگا در تراز۷۰۰هکتوپاسکال در امتداد عرض°۳۷/۵ شمالی از ۱۵ آوریل تا ۱۵ اکتبر. اُمگا برحسب Pa/s بوده و از نشان دادن مقادیر۰/۰۵- تا۰/۰۵+ (Pa/s) صرفنظر شده است. مقادیر اُمگای ۰/۱۰+ و۰/۰۷۵- (Pa/s) به ترتیب با دوره تشکیل و تداوم واچرخند ترکمنستان و ناوه شبه ساکن شرق ترکیه انطباق دارند. پیکان های ضخیم زمان آغاز و پایان تشکیل هر یک از مراکز فشار را نشان می دهند.

۲- واچرخند ترکمنستان مؤلفه‌ی اصلی شکل‌گیری بادهای شمالی بر روی نیمه‌ی شرقی ایران و نیمه‌ی غربی افغانستان و پاکستان است. همانطوریکه شکل ها نشان می دهند، استقرار واچرخند ترکمنستان در محدوده‌ی °۳۷/۵ شمالی و °۵۸ تا °۶۰ طول شرقی(شکل۳-۹e)، به همراه استقرار مرکز کم‌فشار پاکستان در

حد فاصل عرض ۳۰° تا ۳۲/۵° شمالی و طول ۶۵° شرقی(شکل۳-۹f)، یک جریان شمالی پیوسته و شدید را در طول تابستان در محدوده‌ی غرب افغانستان و منتهی‌الیه شرق ایران بدنبال دارد.

۳- بررسی‌های اخیر نشان‌دهنده‌ی آن است که واچرخند ترکمنستان علاوه بر تغییر الگوی گردش ترازهای زیرین جو در بالاسوی منطقه‌ی تشکیل خود و ایجاد ناوه‌ی شبه ایستای شرق ترکیه، به همراه واچرخند شرق مدیترانه، یک موج راسبی بزرگ مقیاس را در جت جنب حاره در پائین سوی منطقه‌ی استقرار خود ایجاد می‌نماید. نتیجه‌ی شکل‌گیری الگوی پشته-ناوه در جت جنب حاره، که به «الگوی جاده‌ی ابریشم»[38] معروف گشته (Enomoto, 2004)، ظهور یک پشته‌ی شبه ساکن در شرق آسیا و در نهایت تشکیل پرفشاری دینامیکی موسم به پرفشار «بونین یا اُگاساوارا»[39] در ماه اوت در ترازهای میانی جو بر روی ژاپن است (Enomoto et al., 2003; Enomoto, 2004).

## ۳-۲-۲-۲- ناوه‌ی شبه ساکن شرق ترکیه

بررسی نقشه‌های ارتفاع و تاوایی تراز ۷۰۰ هکتوپاسکال در مقیاس زمانی پنج روزه از ابتدای ماه آوریل تا پایان ماه اکتبر علاوه بر آشکار ساختن واچرخند ترکمنستان بر جانب شرقی دریای خزر، مبین استقرار یک ناوه‌ی شبه‌ساکن بر جانب غربی دریای خزر می‌باشد(شکل۳-۸ و ۳-۱۱). این ناوه‌ی شبه‌ساکن در یک دوره‌ی زمانی کوتاه، پس از تشکیل و تقویت واچرخند ترکمنستان در نیمه‌ی شرقی ترکیه ظاهر می‌شود (شکل۳-۱۰). با توجه به بررسی نقشه‌ها، زمان ظهور این ناوه بطور متوسط دوره‌ی پنج روزه‌ی دوم تا دوره‌ی پنج روزه‌ی سوم ماه ژوئن(۶-۱۰ تا ۱۵-۱۱ژوئن) می‌باشد(شکل۳-۱۱a و ۳-۱۱b). ناوه‌ی شبه ساکن در حول و حوش ۶ تا ۱۰ ژوئیه به حداکثر عمق خود می‌رسد (شکل۳-۱۱c) و تا اواسط ماه اوت از عمق کافی برخوردار است. از آن پس با ضعیف شدن واچرخند ترکمنستان از عمق ناوه کاسته شده، بطوریکه پس از دوره‌ی پنج روزه‌ی سوم ماه سپتامبر (۱۱-۱۵سپتامبر) جریان در غرب دریای خزر وضعیت تقریباً مداری بخود گرفته و ناوه در شرق ترکیه ناپدید می‌شود(شکل۳-۱۱d).

بررسی متوسط سرعت قائم فصلی(ژوئن، ژوئیه و اوت) در امتداد عرض ۳۷/۵° شمالی بیانگر آن است که استقرار ناوه بر جانب غربی واچرخند ترکمنستان، امکان صعود دائمی هوا را در منطقه‌ای حدفاصل ۴۵° تا ۵۵° طول شرقی فراهم می‌آورد(شکل۳-۹d). با توجه به شکل ۳-۹d صعود هوا (اُمگای منفی)در امتداد طول ۵۰° شرقی به حداکثر میزان خود می رسد. بطوریکه سرعت قائم بالاسو در ترازهای زیرین جو بر جانب

---

38. Silk Road Pattern

39. Bonin High - Ogasawara Anticyclone

غربی دریای خزر به ۰/۱۵- پاسکال بر ثانیه رسیده و صعود تا تراز ۴۰۰ هکتوپاسکال ادامه می یابد. بدین ترتیب در نتیجه بلوکه شدن جریانات غربی و عمیق شده یک ناوه شبه ساکن در غرب واچرخند ترکمنستان، امکان صعود مداوم هوا در تمام تابستان بر روی نیمه‌ی غربی دریای خزر، همچنین منطقه‌ی شمال غرب ایران فراهم می گردد(شکل۳-۱۰). نکته‌ای که در اینجا لازم می‌آید ذکر آن آن است که پیشروی پشته‌ی مستقر بر روی شمال عربستان تا عرض ۳۰° شمالی، بهمراه استقرار واچرخند ترکمنستان بر جانب شرقی ناوه، شکل خاصی به ناوه‌ی شبه‌ساکن در شرق ترکیه می‌دهند(شکل۳-۱۱b و۳-۱۱c). در این رابطه، پیشروی شمال سوی پرفشار عربستان همراه با گسترش واچرخند ترکمنستان، ضمن محدود کردن ناوه، کف آن را مداری نموده و در عین حال، تشکیل دو محور یکی با جهت شمال غربی-جنوب شرقی در شمال غرب ایران و دیگری با جهت شمال شرقی- جنوب غربی در غرب ترکیه را موجب می‌گردد(شکل۳-۱۱c). شکل گیری چنین الگویی، استقرار جانب راست محور ناوه‌ی شرقی را در سراسر تابستان بر روی منطقه ای در حدفاصل شمال غرب ایران تا نیمه‌ی غربی دریای خزر امکانپذیر می نماید که در نهایت تاوایی مثبت وتداوم صعود هوا را برای منطقه فوق بدنبال دارد(شکل۳-۱۰).

با در نظر گرفتن مقادیر متوسط ماهانه تاوایی در مقطع قائم جو در ماه‌های ژوئیه و اوت، حداکثر تاوایی مثبت در امتداد عرض ۳۷/۵° شمالی درحدفاصل ۴۰° تا ۵۰° طول شرقی و در تراز ۸۰۰ هکتوپاسکال بوقوع می پیوندد(شکل۳-۹b و۳-۹c). با توجه به بررسی‌های انجام شده ناوه‌ی شبه ساکن شرق ترکیه از طریق سازوکارهای زیر در وقوع بارش های تابستانه‌ی ایران نقش دارد:

۱- در شرایط نرمال، عمیق شدن ناوه در شرق ترکیه، وقوع بارش های تابستانه در منطقه خزری و بصورت محدودتر مناطق شمال غرب ایران را بواسطه‌ی تداوم تاوایی مثبت و صعود مداوم هوا در ترازهای زیرین و میانی(شکل۳-۹d)در جانب راست محور ناوه، امکانپذیر می سازد.

۲- تقویت واچرخند ترکمنستان و جابجایی شمال سو و گسترش نصف النهاری آن بر جانب شرقی دریای خزر در اغلب موارد موجب عمیق شدن ناوه‌ی شبه‌ساکن بر روی نیمه‌ی غربی ایران می‌گردد. در چنین وضعیتی بخصوص زمانی که ناوه تا شمال خلیج‌فارس پیشروی نماید، افزایش تاوایی مثبت در زیر ناوه، ضمن تقویت زبانه‌ی کم‌فشار تراز زیرین موجب تشکیل کم‌فشارهای کم عمقی در داخل ایران می گردد. کم‌فشارهای مذکور در وقوع بخش عمده‌ی بارش‌های تابستانه‌ی مناطق جنوبی و جنوب شرقی ایران نقشی اساسی دارند. شکل های۳-۱۱e و۳-۱۱f وضعیت سینوپتیکی فوق الذکر را در زمان وقوع بارش روز۱۸ژوئیه۲۰۰۳ در منطقه‌ی جنوب شرقی ایران به ترتیب در تراز ۷۰۰ و۸۵۰ هکتوپاسکال نشان می‌دهند.

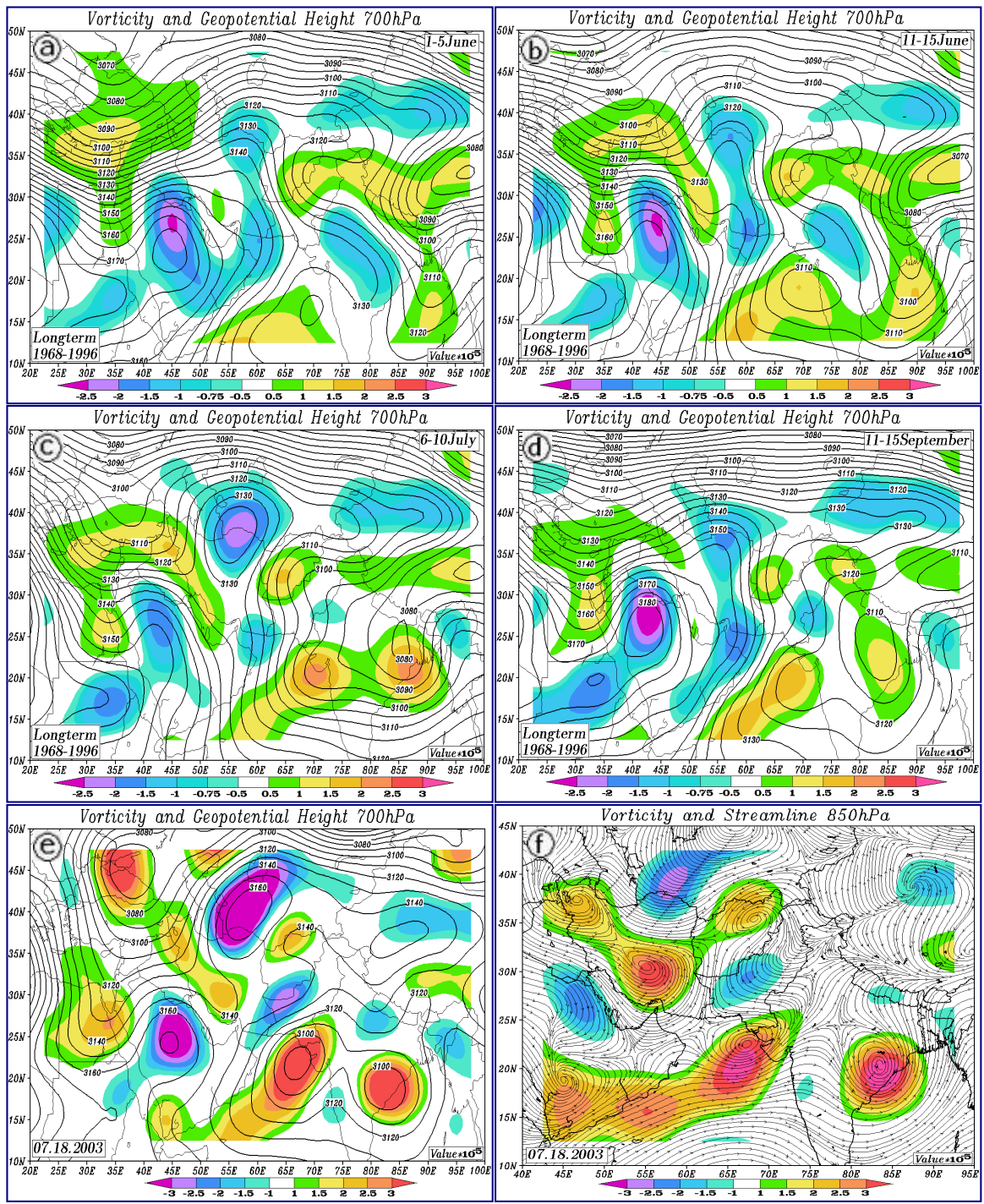

**شکل 3-11.** تشکیل و تکوین ناوه شبه ساکن شرق ترکیه براساس وضعیت متوسط پنج روزه ارتفاع و تاوایی نسبی در تراز ۷۰۰ هکتوپاسکال. (a) دوره پنج روزه قبل از تشکیل ناوه. (b) دوره شکل گیری ناوه همراه با تاوایی مثبت بیشتر از ۱×۱۰$^{-5}$ بر ثانیه. (c) دوره حداکثر عمق و گردش چرخندی. (d) دوره ناپدید شدن ناوه و تضعیف گردش چرخندی در شرق ترکیه. (e) عمیق شدن ناوه شبه ساکن تراز ۷۰۰هکتوپاسکال بر روی نیمه غربی ایران و (f) تشکیل و تقویت مرکز کم فشار جو بر نیمه جنوبی کشور در زمان وقوع بارش۱۸ ژوئیه ۲۰۰۳ در جنوب شرق ایران.

۵۱

### ۳-۲-۲-۳- کم‌فشار پاکستان

بررسی نقشه‌های ارتفاع، خطوط جریان و تاوایی تراز ۸۵۰ هکتوپاسکال در مقیاس زمانی ۵ روزه از آغاز آوریل تا پایان اکتبر نشان می‌دهد که در اواسط ماه می در غرب شبه‌قاره‌ی هند و نیمه‌ی جنوبی پاکستان مرکز کم‌فشاری ظاهر می‌شود. این مرکز کم‌فشار در دوره‌ی گرم سال از ویژگی‌های مشخص و غالب ترازهای زیرین جو در منطقه‌ی جنوب غرب آسیا بشمار می‌رود. مرکز کم‌فشار فوق از ابتدای تشکیل تا پایان ماه ژوئن در محدوده‌ی ۶۸° تا ۷۰° طول شرقی و ۳۰° عرض شمالی جای می‌گیرد(شکل۳-۸a). اما از ابتدای ماه ژوئیه کمی به سمت غرب جابجا شده و در حول و حوش ۶۵° طول شرقی استقرار می‌یابد. با توجه به وضعیت استقرار کم‌فشار در طول فصل تابستان، در متون اقلیمی اصطلاحاً نام «کم‌فشار پاکستان» به آن اطلاق می‌گردد.

در رابطه با علت تشکیل کم‌فشار پاکستان «رمیج»[40] (۱۹۶۶) معتقد است که گرمایش شدید ترازهای زیرین جو به همراه نزول دینامیکی هوا با منشأ منطقه‌ی موسمی غرب هند بر روی منطقه‌ی غرب پاکستان عامل اصلی تشکیل و تقویت این کم‌فشار به شمار می‌رود. از نظر رمیج اگرچه کم‌فشار حرارتی پاکستان بواسطه‌ی آهنگ تابش فصلی در مناطق خشک و بیابانی جنوب غرب آسیا شکل می‌گیرد اما تداوم و تقویت این کم‌فشار در طول دوره گرم سال نتیجه‌ی فرونشینی حاصل از صعود هوای گرم و مرطوب در سیستم موسمی حاکم بر غرب شبه‌قاره‌ی هند می‌باشد. بر پایه‌ی این تئوری، فرونشینی بر بالای کم‌فشار در هماهنگی با بارش‌های موسمی، کم‌فشار حرارتی پاکستان را دچار اوج و حضیض می‌نماید (Ramage,1966; 1971; Flohn *et al.*, 1968). اگرچه برخی از مطالعات، چنین سازوکاری را در رابطه با تقویت دینامیکی کم‌فشار پاکستان مورد تردید قرار داده‌اند(Desai, 1967; Ramamurthi, 1972)، اما تئوری رمیج تنها تئوری موجود در رابطه با تقویت کم‌فشار پاکستان محسوب می‌گردد. در ادامه ویژگی کم‌فشار پاکستان مورد بررسی قرار می‌گیرد.

نقشه‌های ۳-۸a تا ۳-۸c، وضعیت متوسط ماهانه‌ی کم‌فشار پاکستان را در تراز ۸۵۰ هکتوپاسکال به ترتیب در ماه ژوئن، ژوئیه و اوت نشان می‌دهد. همانطوریکه در نقشه‌ها مشاهده می‌شود کم‌فشار پاکستان مؤلفه سینوپتیکی غالب و مسلط ترازهای زیرین جو در منطقه‌ی جنوب-جنوب غرب آسیا به شمار می‌رود. مقایسه‌ی وضعیت کم‌فشار در مقیاس ماهانه بیانگر آن است که کم‌فشار پاکستان در ماه ژوئیه در پی افزایش تاوایی نسبی تا $3 \times 10^{-5}$ بر ثانیه و کاهش ارتفاع ژئوپتانسیل در مرکز کم‌فشار تا ۱۴۱۰ ژئوپتانسیل متر به حداکثر شدت خود می‌رسد(شکل۳-۸b). بررسی‌های اولیه نشاندهنده‌ی آن است که

---

40. Ramage

۵۲

کم‌فشار پاکستان در طی فصل تابستان، بطور متوسط در حدفاصل ۶۵° تا ۶۷° طول شرقی(شکل۱۲a-۳) و ۳۰° عرض شمالی استقرار می‌یابد(شکل۱۲b-۳). بررسی مقادیر متوسط فصلی(ژوئن، ژوئیه و اوت) تاوایی نسبی در مقطع قائم جو در کم‌فشار پاکستان، بیانگر آن است که حداکثر گردش چرخندی در این کم فشار کم عمق در تراز ۸۵۰ هکتوپاسکال بوقوع می‌پیوندد(شکل۱۲a-۳ و ۱۲b-۳). با در نظر گرفتن مقادیر متوسط فصلی تاوایی نسبی، حداکثر متوسط تاوایی در مرکز کم‌فشار به $2/25 \times 10^{-5}$ بر ثانیه بالغ می گردد. همچنین بررسی تاوایی در مقطع قائم جو در مرکز کم فشار نشاندهنده آن است که گردش چرخندی درکم‌فشار پاکستان حداکثر تا تراز ۷۰۰ هکتوپاسکال گسترش یافته و بالاتر از آن گردش واچرخندی گسترده ای جایگزین می‌گردد(شکل۱۲a-۳ و ۱۲b-۳).

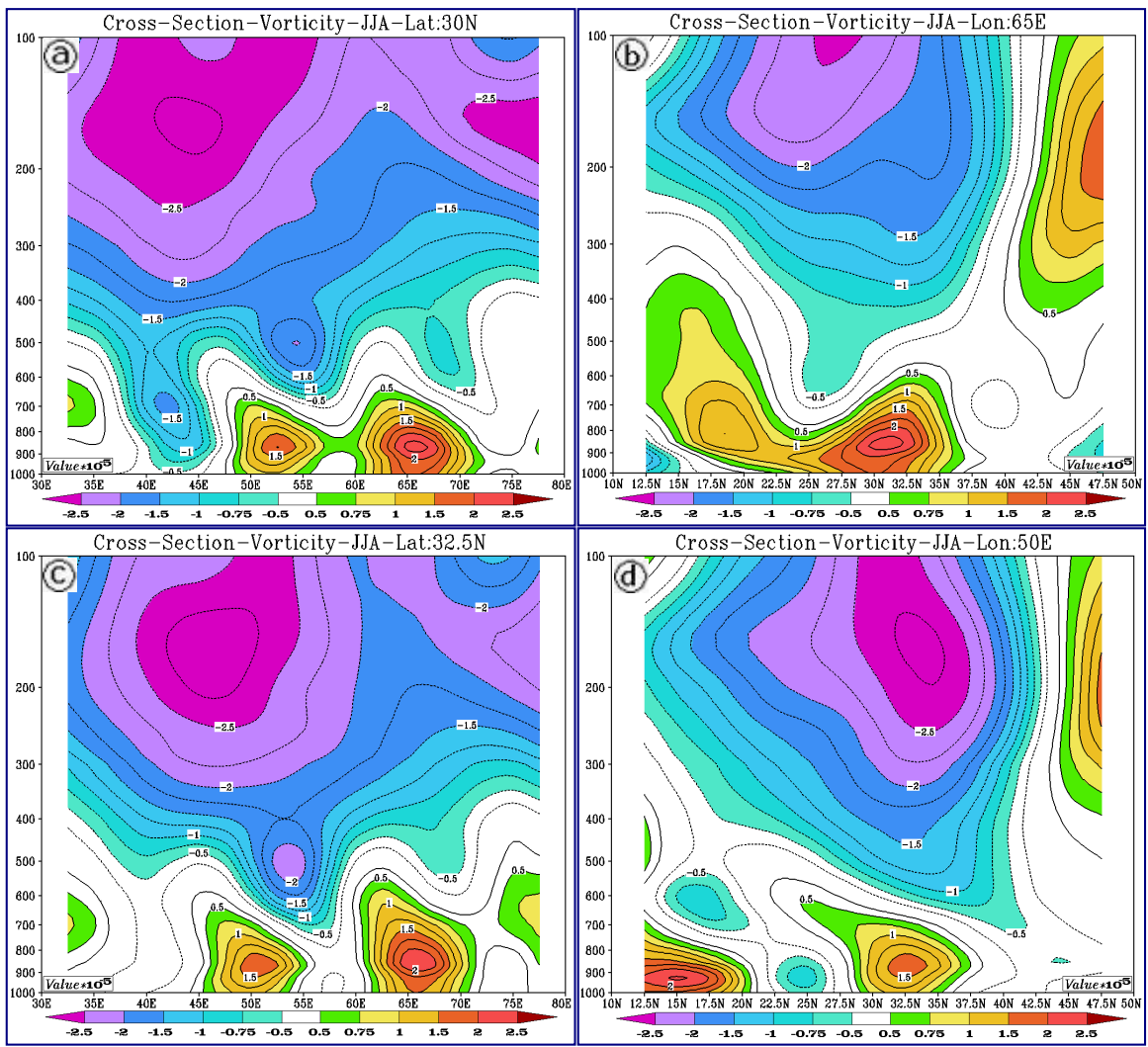

**شکل۱۲-۳.** متوسط فصلی تاوایی نسبی در مقطع قائم جو. (a) و(c) به ترتیب عرض ۳۰° و ۳۲/۵° شمالی. (b) و(d) به ترتیب در امتداد طول ۶۵° و ۵۰° شرقی. تاوایی برحسب $s^{-1}$ $10^{-5}$ است. کلیه داده ها مربوط به یک دوره ۲۹ ساله(۱۹۶۸-۱۹۹۶) می باشد.

۵۳

بررسی ساختار قائم جو در منطقه استقرار کم فشار پاکستان حاکی از آن است که در امتداد عرض ۳۰° شمالی، بواسطه وجود دو زبانه‌ی تاوایی منفی یکی در جانب شرقی کم‌فشار پاکستان و دیگری در مرکز ایران که تا تراز ۷۰۰ هکتوپاسکال پائین آمده‌اند، مرکز کم‌فشار پاکستان را محدود نموده و آن را از سیستم کم‌فشار موسمی در شرق و زبانه کم‌فشار ایران در غرب جدا می‌کنند(شکل۳-۱۲a). تسلط گردش واچرخندی در ترازهای میانی و فوقانی وردسپهر و گسترش پائین سوی زبانه‌های آن تا تراز ۷۰۰ هکتوپاسکال، چنانکه در شکل ۳-۱۲a و ۳-۱۲b مشاهده می‌شود مرکز کم‌فشار پاکستان را از سمت شرق، غرب و جنوب محدود می‌نمایند.

جهت درک میزان صعود هوا در مرکز کم‌فشار پاکستان، شکل ۳-۱۳a مقادیر متوسط فصلی اُمگا را در مقطع قائم جو در امتداد عرض ۳۰° شمالی نشان می‌دهد. با توجه به شکل، حداکثر متوسط صعود هوا در امتداد طول ۶۷° شرقی درحد فاصل ۷۵۰ تا ۸۰۰ هکتوپاسکال بوقوع می‌پیوندند. در این رابطه، حداکثر سرعت قائم بالاسو (اُمگای منفی) در کم‌فشار پاکستان به ۰/۱۱ پاسکال بر ثانیه بالغ می‌گردد.

جهت بررسی تکوین زمانی کم‌فشار پاکستان از نقشه‌های ترکیبی خطوط جریان و تاوایی تراز ۸۵۰ هکتوپاسکال بصورت متوسط ۵ روزه و نمودار هاومولر استفاده شد. بررسی‌ها نشان می‌دهد آستانه‌ی تاوایی $2 \times 10^{-5}$ بر ثانیه در امتداد عرض ۳۰° شمالی در نیمه غربی پاکستان می‌تواند معیار مناسبی جهت بررسی زمانی تکوین کم‌فشار پاکستان و تعیین زمان آغاز و پایان آن باشد(شکل۳-۱۳c). با توجه به بررسی‌های انجام شده، کم‌فشار پاکستان در پایان دهه دوم ماه می در نتیجه‌ی افزایش تاوایی نسبی و پیدایش یک مرکز گردش چرخندی در ۷۰° طول شرقی تشکیل می شود(شکل۳-۱۴c و ۳-۱۳c). کم فشار پاکستان در محدوده‌ی زمانی ۳۰-۲۶ ژوئن تا پایان دهه‌ی دوم ماه اوت به حداکثر عمق و شدت خود می‌رسد و در نهایت در دهه‌ی آخر ماه سپتامبر بدنبال کاهش تاوایی در مرکز کم‌فشار و ناپدید شدن مرکز گردش چرخندی محو می‌گردد(شکل۳-۱۴d ).

شکل۳-۱۳c تکوین زمانی کم‌فشار پاکستان را بر اساس مقادیر تاوایی نسبی از ۱۵ آوریل تا ۱۵ اکتبر در امتداد عرض ۳۰° شمالی نشان می‌دهد. با توجه به نمودار هاومولر و با در نظر گرفتن مقدار تاوایی $2 \times 10^{-5}$ بر ثانیه، زمان آغاز و پایان برای کم‌فشار پاکستان با بررسی نقشه‌های ۵ روزه انطباق دارد. در عین حال نمودار هاومولر بخوبی روند جابجایی غرب سوی کم‌فشار پاکستان را از زمان تشکیل تا زمان ناپدید شدن نشان می‌دهد. بر اساس شکل، کم‌فشار پاکستان در حول و حوش ۷۰° طول شرقی تشکیل می شود(با توجه به تاوایی $2 \times 10^{-5}$، دوره‌ی پنج روزه ۲۰-۱۶ می). اواخر ماه ژوئن بر روی ۶۵° طول شرقی استقرار می یابد و تا پایان دهه‌ی دوم ماه اوت در این محدوده باقی می‌ماند و از آن پس تا پایان ماه سپتامبر در یک جابجایی

۵۴

شرق سوی کند به ۶۷/۵° طول شرقی بازمی گردد(شکل۳-۱۴d و۳-۱۳c).

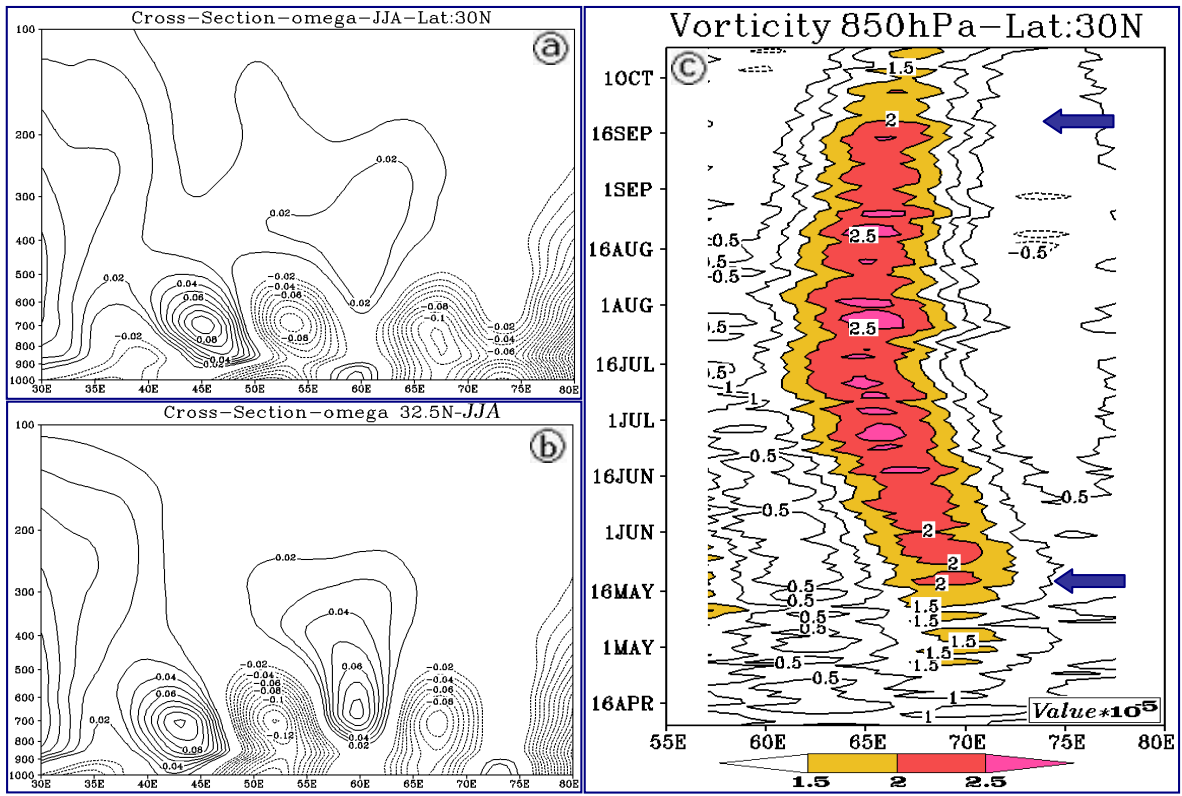

**شکل۳-۱۳.** (a) و(b) متوسط فصلی اُمگا در مقطع قائم جو به ترتیب در امتداد عرض ۳۰° و ۳۲/۵° شمالی. (c)مقادیر متوسط روزانه تاوایی در تراز ۸۵۰ هکتوپاسکال در امتداد عرض ۳۰° شمالی از ۱۵ آوریل تا ۱۵ اکتبر. تاوایی برحسب $10^{-5}$ $s^{-1}$ است. مقادیر تاوایی $2\times10^{-5}$ برثانیه با دوره تشکیل و تداوم کم فشار پاکستان انطباق دارد. کلیه داده ها مربوط به یک دوره ۲۹ ساله(۱۹۶۸-۱۹۹۶) می باشد. پیکان های ضخیم زمان آغاز و پایان تشکیل کم فشار پاکستان را نشان می دهد.

لازم به ذکر است که بررسی تطبیقی مؤلفه‌های گردش جو در منطقه‌ی جنوب غرب آسیا وجود یک همزمانی در آغاز و پایان شکل‌گیری مؤلفه‌ها را آشکار می‌سازد. بدین ترتیب که تغییر ناگهانی در موقعیت و شدت جریان جت جنب حاره و جت شرقی در ترازهای فوقانی (شکل۳-6d) ، با زمان ظهور جت سومالی، واچرخند ترکمنستان (شکل۳-۱۰) و کم‌فشار پاکستان(شکل۳-۱۳c) در ترازهای زیرین جو بخوبی انطباق دارد. از سوئی همزمانی تشکیل و تکوین واچرخند ترکمنستان و کم‌فشار پاکستان، شکل‌گیری و تداوم باد ۱۲۰روزه‌ی سیستان را بطور مناسبی تبیین می‌کند. بر این اساس باد ۱۲۰ روزه‌ی سیستان صرفاً یک باد حرارتی ناشی از اختلافات حرارتی محلی نبوده، بلکه بواسطه شرایط سینوپتیکی حاکم بر نیمه‌ی شرقی ایران و در نتیجه‌ی واداشت‌های دینامیکی شکل گرفته و تداوم می‌یابد. شکل‌های

3-9e و 3-9f الگوی متوسط فصلی جریان هوا و نحوه وقوع باد120روزه را در ترازهای زیرین جو بر روی نیمه‌ی شرقی ایران نشان می‌دهد.

### 4-2-2-3- زبانه‌ی کم‌فشار زاگرس

بر اساس برخی تحقیقات انجام شده ویژگی سینوپتیکی غالب در ترازهای زیرین جو بر روی خاورمیانه در فصل تابستان، ناوه‌ی حرارتی کم‌ضخامتی موسوم به «ناوه‌ی فشاری خلیج‌فارس»[41] می‌باشد ( Bitan and Saaroni, 1992; Saaroni and Ziv, 2000). با توجه به تحقیقات فوق، ناوه‌ی فشاری خلیج‌فارس ناشی از گسترش غرب سوی موسمی جنوب آسیاست که پس از عبور از خلیج‌فارس تا مناطق جنوبی ترکیه و دریای اژه گسترش یافته و بعنوان یک ویژگی پایدار در نقشه‌های متوسط ماهانه و فصلی در سطح دریا مشاهده می‌گردد(Bitan and Saaroni, 1992). تداوم ناوه‌ی فشاری خلیج‌فارس یکی از مؤلفه‌های اصلی شکل‌گیری«باد شمالی»[42] در طول فصل تابستان در شرق مدیترانه دانسته شده است ( Saaroni and Ziv, 2000; Zaitchik et al., 2007). بررسی سازوکار وقوع بارش‌های تابستانه‌ی ایران بیانگر آن است که تشکیل و تقویت مراکز چرخندی و کم‌فشارها بر روی زبانه‌های کم‌فشار در نیمه جنوبی کشور، علت اصلی وقوع بارش‌های تابستانه‌ی جنوب و جنوب شرق کشور می‌باشد. با توجه به اینکه تاکنون در متون اقلیمی کشور این مؤلفه‌ی سینوپتیکی مورد توجه قرار نگرفته، در ادامه خصوصیات زبانه‌ی کم‌فشار مورد بررسی قرار می‌گیرد.

بررسی نقشه‌های متوسط پنج روزه ارتفاع و تاوایی تراز 850 هکتوپاسکال از ابتدای آوریل تا پایان اکتبر نشاندهنده‌ی آن است که در حول و حوش دوره‌ی پنج روزه‌ی اول ماه ژوئن در نتیجه‌ی افزایش تاوایی مثبت در عرض°30 شمالی و بر روی زاگرس، زبانه‌ی کم‌فشاری بر روی غرب ایران شکل می‌گیرد(شکل3-14a). در طی ماه ژوئن، مقادیر تاوایی بر روی زاگرس (محور اصلی گسترش زبانه) به $2/5 \times 10^{-5}$ بر ثانیه می‌رسد. در این زمان همان‌طوری‌که نقشه‌ی متوسط ماه ژوئن(شکل3-8a) نشان می‌دهد، زبانه‌ی کم‌فشار در امتداد رشته کوه زاگرس جهتی شمال غربی-جنوب شرقی پیدا نموده و تا شرق مدیترانه گسترش می‌یابد. در طی ماه ژوئیه و اوت حداکثر متوسط تاوایی مثبت به میزان $2 \times 10^{-5}$ بر ثانیه بر روی زاگرس مشاهده می‌شود (شکل3-8b و3-8c). زبانه‌ی کم‌فشار فوق، که از این پس آن را «زبانه‌ی کم‌فشار زاگرس» می‌نامیم تا دهه‌ی سوم ماه سپتامبر در امتداد رشته‌ی کوه زاگرس تداوم می‌یابد(شکل 3-14b) و

---

41. Persian Gulf Trough

42. Etesian Wind

56

از آن پس بدنبال کاهش تاوایی مثبت و افزایش فشار در غرب ایران ناپدید می‌گردد. تبعیت زبانه‌ی کم‌فشار از جهت استقرار رشته کوه زاگرس و تداوم منطقه‌ی حداکثر تاوایی مثبت بر روی منطقه‌ی مرتفع زاگرس نشان‌دهنده نقش واداشت‌های حرارتی این رشته کوه در تشکیل و تداوم این زبانه‌ی کم‌فشار در نیمه‌ی غربی ایران می‌باشد. به منظور بررسی این تئوری و تعیین میزان نقش رشته کوه زاگرس در تشکیل و تکوین زبانه‌ی کم فشار ایران، گرمایش درو، گرمایش دررو در ترازهای ۸۵۰ و ۷۰۰ هکتوپاسکال بر روی جنوب غرب آسیا با استفاده از «معادله انرژی ترمودینامیک» برای ماه ژوئیه محاسبه شد. شکل‌های ۳-۱۵a و ۳-۱۵b به ترتیب میزان گرمایش درو و فرارفت قائم گرما را به کلوین در روز در تراز ۷۰۰ هکتوپاسکال نشان می‌دهند. با توجه به شکل ۳-۱۵a نیمه غربی ایران در ماه ژوئیه حداکثر گرمایش در رویی به میزان ۱/۴ کلوین در روز را تجربه می‌کند. نکته جالب اینکه گرمایش دررویی ایجاد شده در تراز ۷۰۰ هکتوپاسکال انطباق خوبی با محور استقرار رشته کوه زاگرس دارد. بررسی شکل ۳-۱۵b مربوط به فرارفت قائم گرما بیانگر آن است که بیشینه گرمای ایجاد شده بر روی نیمه غربی ایران در ترازهای ۸۵۰ و ۷۰۰ هکتوپاسکال، ناشی از فرارفت قائم گرما از روی رشته کوه مرتفع زاگرس می‌باشد. بطوریکه بیشینه فرارفت قائم گرما در محدوده ۳۲° تا ۳۵° عرض شمالی و ۵۲° طول شرقی به ۱/۴ کلوین در روز بالغ می‌گردد. نتیجه اینکه، تشکیل و گسترش زبانه کم فشار بر روی نیمه غربی ایران، ناشی از واداشت‌های حرارتی زاگرس بعنوان یک منبع گرمای محسوس ارتفاع یافته بر روی غرب ایران می‌باشد. شایان ذکر است که کم فشار پاکستان نیز با توجه به شکل ۳-۱۵a و ۳-۱۵b از الگویی مشابه کم فشار ایران جهت تشکیل و تقویت خود بهره می‌برد. یافته‌های فوق با نتایج بررسی‌های پیشین مبنی بر تشکیل زبانه‌ی کم فشار بر روی خاورمیانه در نتیجه گسترش غرب سوی موسمی جنوب آسیا بر روی جنوب غرب آسیا تفاوت دارد (Bitan and Saaroni, 1992). در واقع صحیح‌تر آن است که نام این ناوه فشاری کم عمق را «زبانه‌ی کم فشار زاگرس» نامگذاری نمائیم.

بررسی مقادیر متوسط فصلی تاوایی در مقطع قائم جو نشان‌دهنده‌ی آن است که حداکثر گردش چرخندی در زبانه کم فشار بطور متوسط در طول ۵۰° شرقی و عرض ۳۲/۵° شمالی در تراز ۸۵۰ هکتوپاسکال بوقوع می‌پیوندد (شکل ۳-۱۲c و ۳-۱۲d). با توجه به موقعیت مذکور، حداکثر گردش چرخندی در زبانه‌ی کم‌فشار بر روی مناطق مرتفع زاگرس اتفاق می‌افتد و در بالای تراز ۸۵۰ هکتوپاسکال میزان تاوایی مثبت به سرعت کاهش می‌یابد و گردش چرخندی حداکثر تا تراز ۷۰۰ هکتوپاسکال مشاهده می‌گردد (شکل ۳-۱۲c). بالای تراز فوق، بواسطه‌ی استقرار پرفشار بر روی مرکز ایران گردش واچرخندی جایگزین می‌گردد.

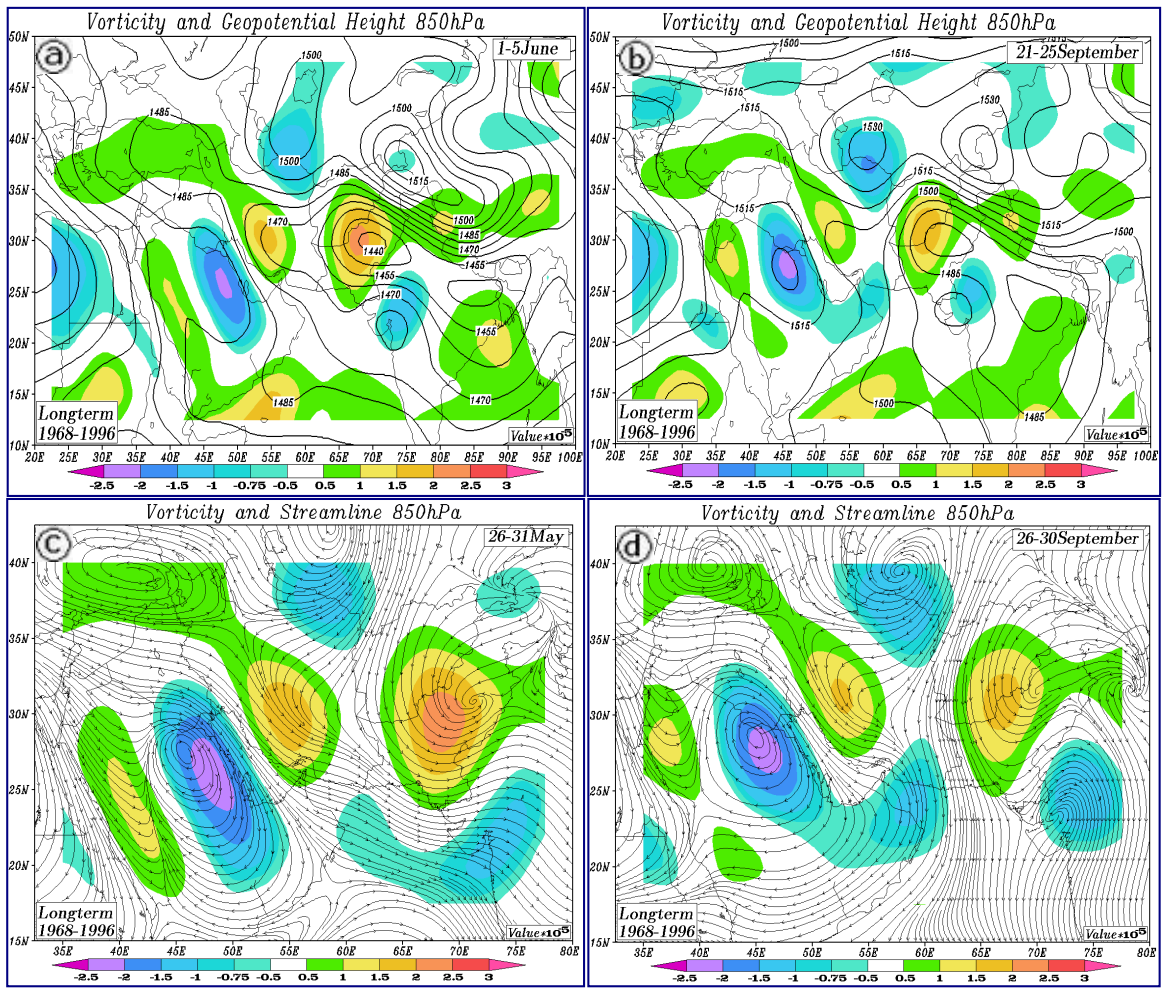

**شکل۳-۱۴.** (a) و(b) دوره های پنج روزه آغاز و پایان تشکیل زبانه ی کم فشار زاگرس. (c) و(d) به ترتیب دوره های پنج روزه آغاز(c) و پایان(d) تشکیل کم فشار پاکستان.

جهت بررسی میزان صعود هوا در زبانه‌ی کم‌فشار، متوسط سرعت قائم نیز برای فصل تابستان در دو عرض °۳۰ (عرض ترجیحی ماه ژوئن) و °۳۲/۵ شمالی(عرض ترجیحی ماه‌های ژوئیه و اوت) مورد مطالعه قرار گرفت(شکل ۳-۱۳a و۱۳b-۳). با توجه به شکل‌ها، حداکثر صعود هوا در حول و حوش طول °۵۲/۵ شرقی در تراز ۷۰۰ هکتوپاسکال بوقوع می‌پیوندند. در این رابطه، میزان متوسط سرعت قائم در عرض‌های °۳۰ و °۳۲/۵ شمالی به ترتیب ۰/۰۹ و۰/۱۳ پاسکال برثانیه می‌باشد(شکل۳-۱۳a و۱۳b-۳).

شکل۱۵c-۳ تکوین زمانی زبانه‌ی کم‌فشار ایران را در امتداد عرض °۳۲/۵ شمالی نشان می‌دهد. در این نمودار مقادیر تاوایی $1/5 \times 10^{-5}$ بر ثانیه در محدوده‌ی °۵۰ تا °۵۵ طول شرقی، معیار مناسبی جهت بررسی تکوین زبانه‌ی کم‌فشار در نیمه‌ی غربی ایران می‌باشد. با توجه به مقادیر تاوایی $1/5 \times 10^{-5}$ بر ثانیه، زبانه‌ی کم‌فشار در روزهای آغازین ماه ژوئن در °۵۲/۵ طول شرقی ظاهر شده و تا پایان دهه‌ی دوم سپتامبر در

حدفاصل ۵۰° تا ۵۲/۵° طول شرقی تداوم می‌یابد. با در نظر گرفتن شکل۳-۱۵c میزان شدت، همچنین مدت تداوم زبانه‌ی کم‌فشار ایران در مقایسه با کم‌فشار پاکستان (شکل ۳-۱۳c) کمتر و کوتاه‌تر است.

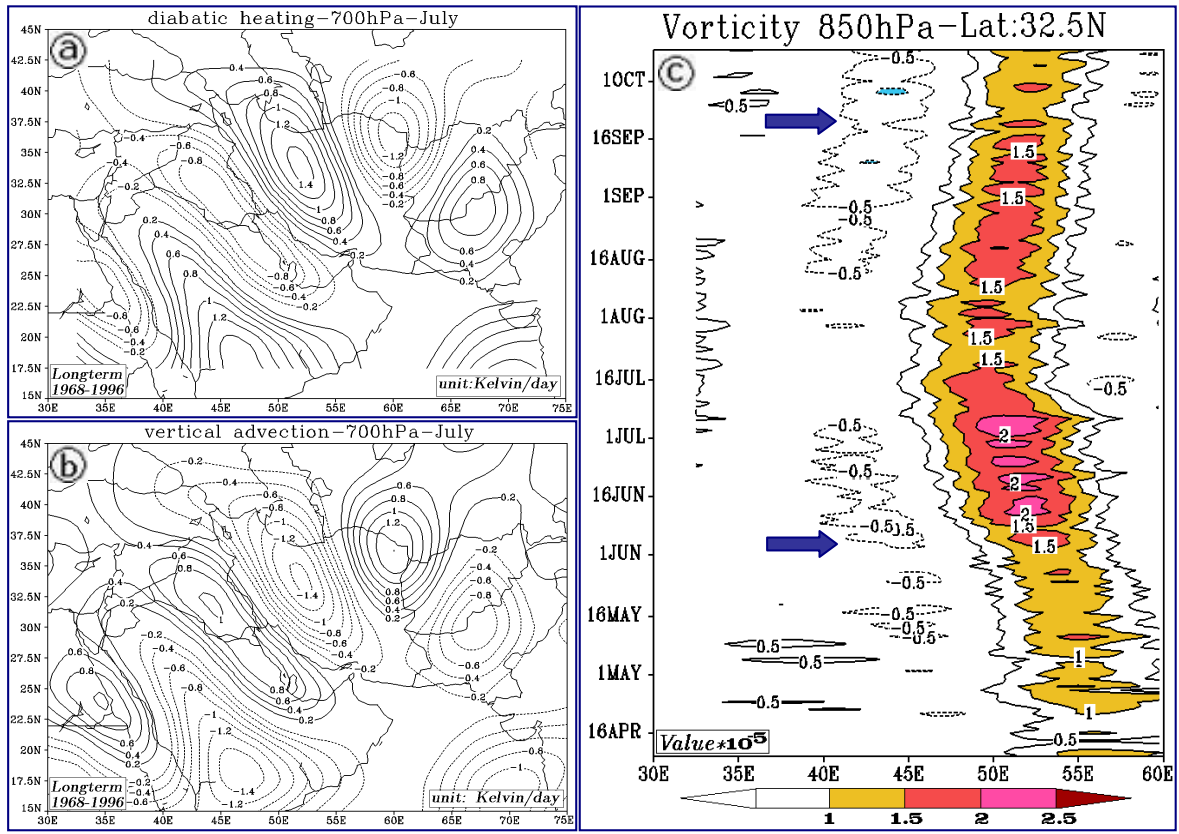

**شکل۳-۱۵.** (a) و(b) به ترتیب گرمایش در رو و فرارفت قائم گرما در ماه ژوئیه در تراز ۷۰۰هکتوپاسکال. (c) نمودار هاوم‌ولر مقادیر متوسط روزانه‌ی تاوایی نسبی از ۱۵ آوریل تا ۱۵ اکتبر در تراز ۸۵۰ هکتوپاسکال و در امتداد عرض ۳۲/۵° شمالی. پیکان‌های ضخیم زمان آغاز و پایان تشکیل زبانه‌ی کم فشار زاگرس را نشان می‌دهند.

## ۳-۲-۳ جریان هوا در ترازهای زیرین جو بر روی جنوب غرب آسیا

بررسی جریان هوا در ترازهای زیرین جو بر روی جنوب غرب آسیا بیانگر ارتباط بین بادهای منطقه ای موسوم به «باد ۱۲۰ روزه سیستان» و «باد شمال» با نحوه استقرار مراکز فشار در وردسپهر زیرین است. با توجه به بررسی های انجام شده، باد ۱۲۰ روزه سیستان صرفاً یک باد محلی و حرارتی نبوده بلکه نتیجه شکل گیری و تداوم مرکز کم فشار پاکستان و واچرخند ترکمنستان به ترتیب در شرق و شمال شرق فلات ایران می باشد. در مقابل، وزش باد شمال در منطقه خلیج فارس نیز برخلاف تصور موجود، منشأ دینامیکی داشته و به دنبال شکل گیری زبانه ی کم فشار زاگرس بر نیمه غربی ایران و جابجایی شمال غرب سوی مرکز پرفشار عربستان و استقرار مداوم آن در بالای عرض ۲۷/۵° شمالی در طول فصل تابستان بر نیمه شمالی

خلیج فارس ظاهر می شود(شکل۳-۱۶).

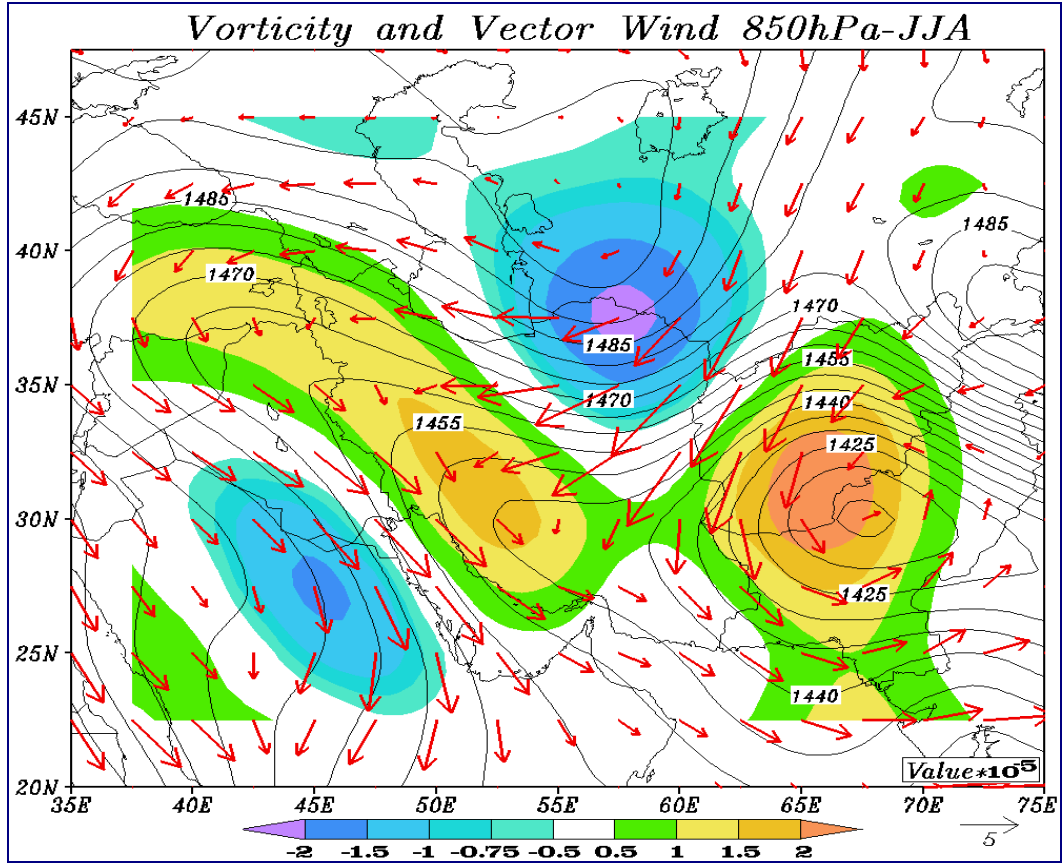

**شکل۳-۱۶.** وضعیت متوسط ارتفاع و تاوایی نسبی به همراه جهت و شدت وزش بادها در تراز ۸۵۰ هکتوپاسکال برای یک دوره ۲۹ ساله(۱۹۹۶-۱۹۶۸) بر روی جنوب غرب آسیا برای فصل تابستان. تاوایی برحسب $s^{-1}$ ۱۰$^{-5}$ ، ارتفاع ژئوپتانسیل برحسب ژئوپتانسیل متر و سرعت باد برحسب متر بر ثانیه می باشد.

شکل۳-۱۶ جهت و شدت متوسط وزش باد را به همراه مقادیر متوسط تاوایی و ارتفاع ژئوپتانسیل در تراز ۸۵۰ هکتوپاسکال بر روی جنوب غرب آسیا برای فصل تابستان نشان می دهد. با توجه به شکل، استقرار مداوم و ممتد واچرخند ترکمنستان و کم فشار پاکستان به ترتیب در شمال شرق و حد شرقی فلات ایران موجب شکل گیری یک جریان شمالی مداوم و نسبتاً قوی از مرکز واچرخند ترکمنستان به سمت مرکز کم فشار پاکستان و همچنین به سمت زبانه کم فشار زاگرس در داخل ایران می گردد. در این میان قویترین بادها بخاطر شیب فشار و شیب تاوایی زیاد و فاصله کم بین کم فشار پاکستان و واچرخند ترکمنستان در منتهی الیه شرق ایران مشاهده می شود(شکل۳-۱۶).

با توجه به شکل های ۳-۱۷a و ۳-۱۷c این باد شمالی(باد۱۲۰ روزه سیستان) در طول ۶۲/۵° شرقی و حدفاصل عرض ۳۲/۵° تا ۳۵° شمالی در تراز ۸۵۰ هکتوپاسکال به حداکثر شدت خود در مقیاس فصلی

۶۰

دست می یابد. در این رابطه حداکثر سرعت متوسط باد شمالی(باد ۱۲۰ روزه سیستان) در حول و حوش عرض ۳۴° شمالی به ۲۸ کیلومتر در ساعت بالغ می گردد(شکل ۱۷c-۳).

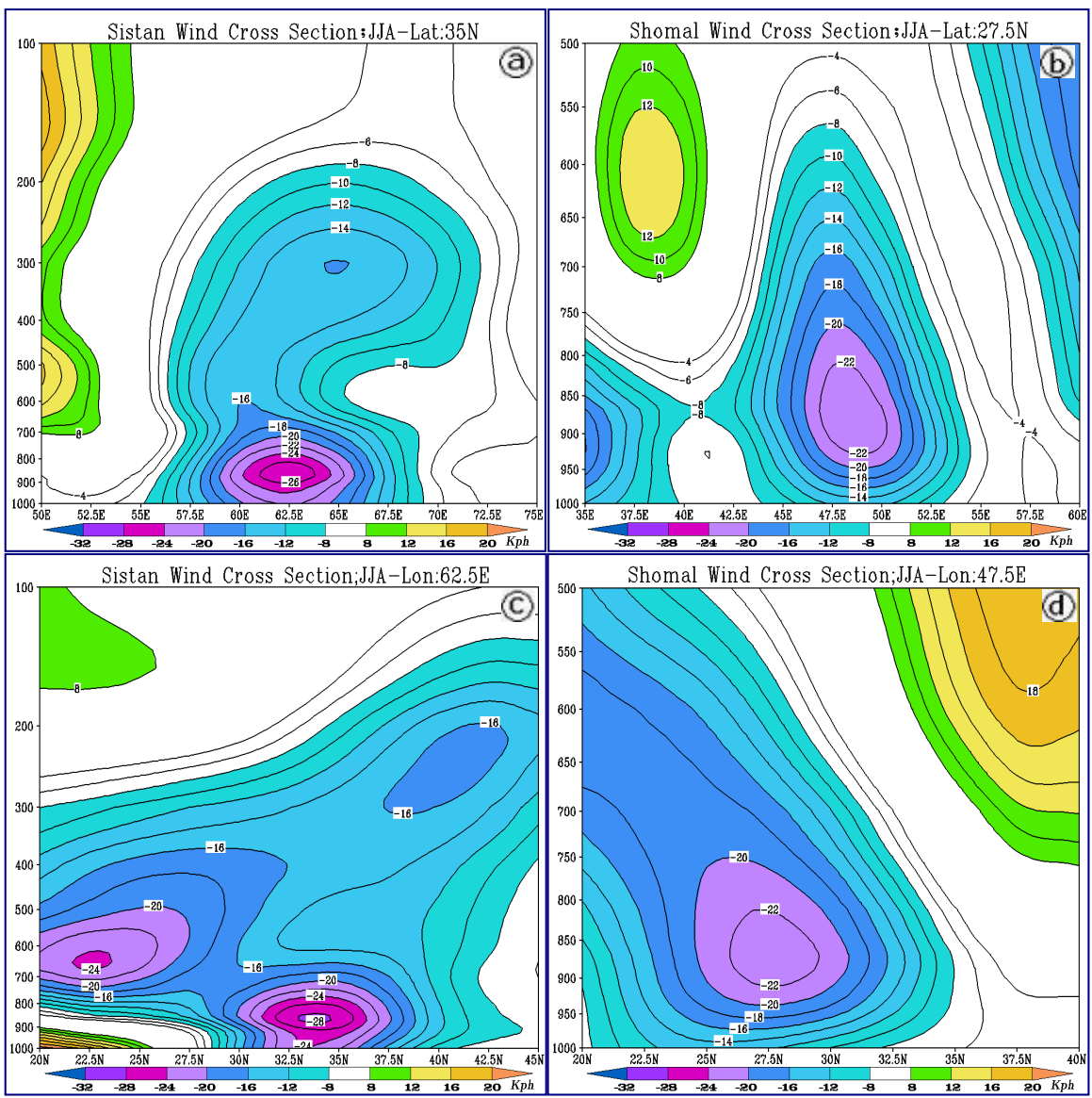

**شکل۱۷-۳.** وضعیت متوسط فصلی مؤلفه باد نصف النهاری در مقطع قائم جو برای یک دوره ۲۹ ساله(۱۹۹۶-۱۹۶۸). (a) و(c) به ترتیب در امتداد عرض ۳۵° شمالی وطول ۶۲/۵° شرقی برای باد سیستان. (b) و(d) به ترتیب در امتداد عرض ۲۷/۵° شمالی و طول ۴۷/۵° شرقی برای باد شمال. سرعت باد برحسب کیلومتر در ساعت می باشد.

بررسی تکوین زمانی باد ۱۲۰ روزه سیستان بخوبی با زمان تشکیل و ناپدید شدن مرکز کم فشار پاکستان و واچرخند ترکمنستان انطباق دارد. با در نظر گرفتن شکل ۱۸a-۳ باد شمالی موسوم به باد ۱۲۰ روزه به دنبال تشکیل مراکز فشار فوق در اواسط ماه می بر روی منتهی الیه شرق ایران ظاهر شده و تا اواسط ماه

اکتبر تداوم می یابد. باد ۱۲۰ روزه سیستان در حدفاصل اواسط ژوئن تا روزهای آغازین ماه سپتامبر به حداکثر شدت و وسعت منطقه ای خود دست می یابد و در همین دوره در طی روزهای میانی ماه ژوئیه حداکثر سرعت خود را تجربه می کند. در این زمان بیشینه سرعت باد در تراز ۸۵۰ هکتوپاسکال به ۳۲ کیلومتر در ساعت بالغ می گردد(شکل ۳-۱۸a). نکته دیگر اینکه با توجه به دینامیکی بودن واچرخند ترکمنستان و ضخامت زیاد آن، بادهای شمالی منتهی الیه شرق ایران تنها مختص ترازهای زیرین جو نبوده بلکه تا تراز ۲۰۰ هکتوپاسکال نیز قابل مشاهده هستند(شکل ۳-۱۷a و ۳-۱۷c). بدین ترتیب می توان چنین نتیجه گیری کرد که باد معروف ۱۲۰ روزه سیستان یک جریان هوا با منشأ دینامیکی است که ظهور، تقویت و تضعیف و ناپدید شدن آن وابسته به وضعیت حاکم بر مراکز فشار ایجاد کننده آن یعنی واچرخند ترکمنستان و کم فشار پاکستان می باشد.

در مقابل «باد شمال» زمانی ظاهر می گردد که مرکز پرفشار دینامیکی عربستان و زبانه کم فشار زاگرس به ترتیب بر جانب غربی و شرقی خلیج فارس استقرار یابند(شکل۳-۱۶). با در نظر گرفتن نحوه قرارگیری مراکز فشار فوق، باد شمال عمدتاً از یک جهت وزش غالب شمال غربی برخوردار می باشد(شکل۳-۱۶). بررسی ها نشان می دهد باد شمال نیز همچون باد۱۲۰ روزه سیستان در حول و حوش اواسط ماه می در محدوده شمال غربی خلیج فارس ظاهر شده و به تبعیت از زبانه کم فشار زاگرس و پرفشار عربستان تا دهه اول ماه اکتبر تداوم می یابد(شکل ۳-۱۸b). با توجه به شکل ۳-۱۸b و مقایسه آن با ۳-۱۸a می توان گفت که باد شمال در مقایسه با باد ۱۲۰ روزه از تغییرات قابل ملاحظه ای در امتداد طول جغرافیایی(امتداد مداری) برخوردار است. به نظر می رسد علت اصلی تغییرات طولی(مداری) وزش باد شمال ناشی از جابجایی مکانی مرکز پرفشار عربستان در طول فصل تابستان باشد.

باد شمال همانطوریکه شکل های ۳-۱۷b و ۳-۱۷d نشان می دهند در تراز ۹۰۰ تا ۸۵۰ هکتوپاسکال به حداکثر شدت خود می رسد. حداکثر سرعت متوسط باد شمال در تراز فوق در حدفاصل °۴۷/۵ تا °۵۰ طول شرقی و عرض °۲۷/۵ شمالی به ۲۲ کیلومتر در ساعت بالغ می گردد که در مقایسه با حداکثر سرعت متوسط باد ۱۲۰ روزه حدود ۶ کیلومتر در ساعت ضعیف تر است(شکل های۳-۱۷a و ۳-۱۷c را با ۳-۱۷b و ۳-۱۷d مقایسه کنید).

باد شمال در حدفاصل اواسط ماه ژوئن تا اواسط ماه ژوئیه به حداکثر شدت وزش خود می رسد. در این دوره و در روزهای پایانی ماه ژوئن حداکثر مطلقی حدود ۲۸ کیلومتر در ساعت را در مقیاس اقلیمی کسب می کند(شکل ۳-۱۸b). نکته دیگر اینکه باد شمال نیز همچون باد ۱۲۰ روزه سیستان از ضخامت

قابل ملاحظه ای برخوردار می باشد. بطوریکه وزش آن تا حول و حوش تراز ۵۰۰ هکتوپاسکال قابل مشاهده است(شکل ۳-۱۷b و ۳-۱۷d).

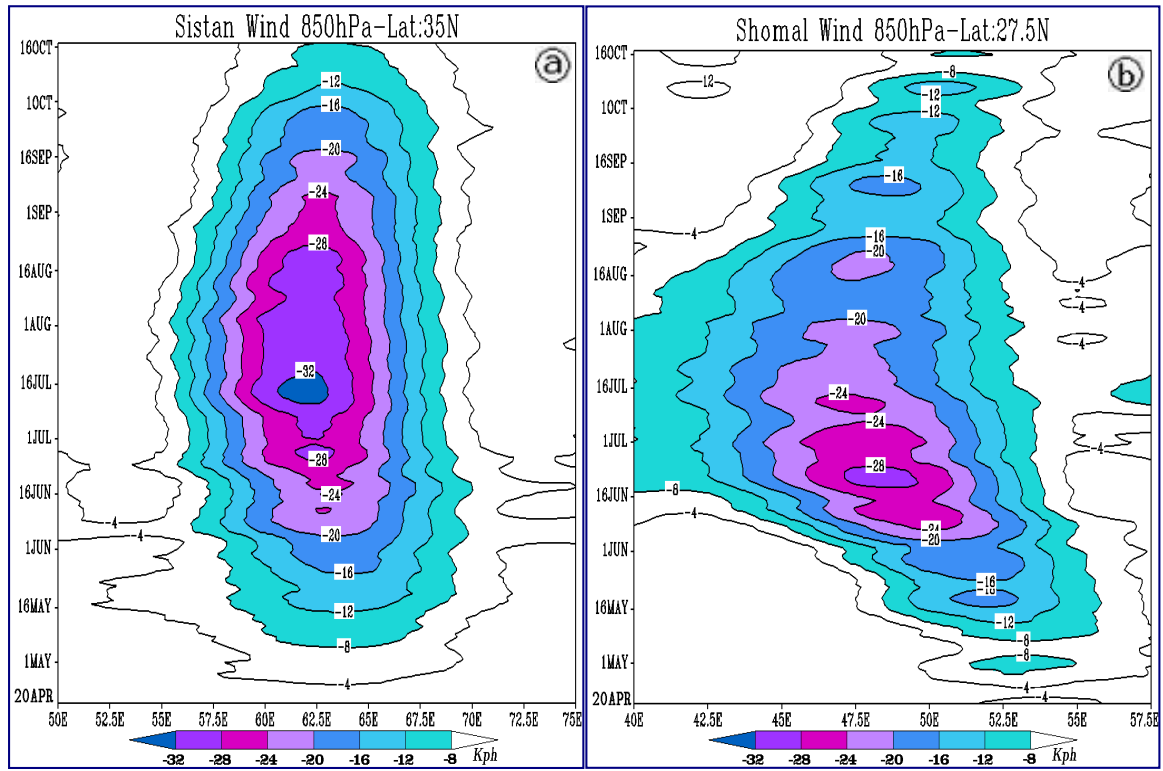

**شکل ۳-۱۸.** وضعیت متوسط روزانه مؤلفه باد نصف النهاری برای یک دوره ۲۹ ساله(۱۹۶۸-۱۹۹۶). (a) در امتداد عرض˚۳۵ شمالی برای باد سیستان. (b) در امتداد عرض˚۲۷/۵ شمالی برای باد شمال. سرعت باد برحسب کیلومتر در ساعت می باشد.

## ۳-۳ خلاصه نتایج

در پاسخ به سؤال اول تحقیق مبنی بر اینکه« گردش جو تابستانه بر روی جنوب غرب آسیا از چه ساختاری برخوردار است و هر یک از مؤلفه های آن دارای چه ویژگی هایی است؟» مهمترین نتایج بشرح زیر می باشد:

• بررسی مرکز پرفشار جنب حاره در تراز فوقانی بیانگر آن است که در اوج تابستان در طی دوره‌ی ۵ روزه نخست ماه ژوئیه و بصورت کاملاً ناگهانی مرکز پرفشاری بر روی جنوب ایران تشکیل شده و واچرخند بزرگ مستقر بر روی جنوب و جنوب غرب آسیا به شکل دو سلولی در می‌آید. این مرکز پرفشار تا پایان دهه دوم ماه اوت بر روی ایران استقرار یافته و سپس در طی دوره‌ی ۵ روزه‌ی ۲۱ تا ۲۵ اوت در طی یک جابجایی شرق سوی ناگهانی، مجدداً مرکز اصلی پرفشار بر جانب جنوبی فلات تبت جای می گیرد.

• بررسی وردش‌های زمانی و مکانی پرفشارهای جنب حاره تراز میانی بر روی جنوب غرب آسیا نشان داد که پرفشار عربستان از ابتدای آوریل تا پایان ماه می جابجایی شمال غرب سویی را به سمت شمال شبه جزیره‌ی عربستان تجربه می‌کند و در حد فاصل دوره‌ی پنج روزه‌ی پایان ماه می تا دوره‌ی پنج روزه‌ی پایان ماه ژوئن، حداکثر تاوایی منفی و بالاترین مقادیر ارتفاع ژئوپتانسیل خود را کسب می نماید و پس از آن در طی ماه ژوئیه و اوت از شدت گردش واچرخندی آن کاسته شده و ارتفاع ژئوپتانسیل درکنتور مرکزی آن کاهش می یابد. در پنج روز پایانی ماه ژوئن، سلول پرفشار دیگری در امتداد شمال شرقی پرفشار عربستان در شمال خلیج فارس بر روی ایران بسته می‌شود. این مرکز پرفشار که آن را «پرفشار ایران» می‌نامیم از اواسط ماه ژوئیه تا پایان دهه‌ی دوم ماه اوت به حداکثر شدت و گستردگی خود می‌رسد. در طی این دوره، بالاترین مقادیر تاوایی منفی و ارتفاع ژئوپتانسیل در منطقه‌ی جنوب غرب آسیا بر روی ایران مشاهده می‌شود.

• پرفشار ایران در تراز میانی از زمان تشکیل خود در روزهای پایانی ماه ژوئن تا ناپدید شدن آن در دوره‌ی پنج روزه‌ی پایان ماه اوت، بطور متوسط در طول ۵۰° شرقی و عرض ۳۲/۵° شمالی استقرار یافته و حداکثر تاوایی منفی آن در طول ۵۲/۵° شرقی مشاهده می گردد. بررسی تطبیقی مقادیر ارتفاع ژئوپتانسیل و تاوایی نسبی در سلول پرفشار ایران نشاندهنده آن است که مقادیر تاوایی $-2/5 \times 10^{-5}$ بر ثانیه می تواند معیار مناسبی جهت تشخیص تکوین زمانی و مکانی پرفشار ایران در تراز ۵۰۰ هکتوپاسکال باشد.

• بررسی وردش های زمانی و مکانی جت جنب حاره بر روی جنوب غرب آسیا نشان داد که جت از ابتدای آوریل تا پایان دهه اول ماه می جابجایی چندانی از خود نشان نمی‌دهد و در عرض ۲۵° تا ۲۷° شمالی استقرار می‌یابد. اما در حد فاصل دهه‌ی دوم ماه می تا دهه اول ماه ژوئن، جت بر روی جنوب غرب آسیا به یکباره حدود ۸ درجه به سمت شمال جابجا شده و در موقعیت تابستانی خود قرار می‌گیرد. در ماه ژوئیه با شکل‌گیری مرکز پرفشار جنب حاره بر روی ایران، جت جنب حاره جنوب غرب آسیا به شمالی‌ترین موقعیت خود در طول سال منتقل شده و بر جانب شمالی دریای خزر جای می‌گیرد.

• نتایج بررسی ها نشان داد که استقرار رژیم گردش بزرگ مقیاس تابستانه بر روی جنوب غرب آسیا با جابجایی ناگهانی شمال سوی جت جنب حاره، همچنین با پدیدار شدن و تقویت ناگهانی جت شرقی در تراز فوقانی و جت سومالی در تراز زیرین در دوره‌ی پنج روزه‌ی دوم ماه ژوئن (۶ تا ۱۰ژوئن) همراه است.

• نتایج بررسی ها نشان داد که مؤلفه‌های اصلی گردش جو در ترازهای زیرین جو بر روی جنوب غرب آسیا عبارتند از: ۱) واچرخند ترکمنستان. ۲) ناوه‌ی شبه ساکن شرق ترکیه. ۳) کم‌فشار پاکستان. ۴)زبانه‌ی

کم‌فشار زاگرس. ۵) پرفشار عربستان. مجموع مؤلفه های فوق ویژگی های اقلیمی غالب منطقه جنوب غرب آسیا را در طول دوره گرم سال توجیه می نماید.

• نتایج بررسی نشان داد که واچرخند ترکمنستان در تراز ۷۰۰ هکتوپاسکال ابتدا در طول ماه ژوئن بصورت یک پشته بر جانب شرقی خزر با حداکثر تاوایی منفی $-2 \times 10^{-5}$ بر ثانیه ظاهر شده و سپس در طی ماه ژوئیه و اوت بصورت یک واچرخند بسته در عرض ۳۷/۵° شمالی و در حدفاصل ۵۷° تا ۶۰° طول شرقی استقرار می‌یابد. همچنین نتایج تحقیق بیانگر آن است که نزول هوا در این واچرخند از بالای تراز ۲۰۰ هکتوپاسکال آغاز شده و تا تراز ۹۰۰ هکتوپاسکال ادامه می‌یابد. حداکثر نزول هوا نیز در تراز ۷۰۰ هکتوپاسکال بوقوع می‌پیوندند. بررسی‌های انجام شده نشان داد که واچرخند ترکمنستان با توجه به موارد زیر گردش جو بر روی جنوب غرب آسیا را از خود متأثر می‌سازد: ۱) عامل اصلی شکل گیری و تداوم ناوه شبه ساکن شرق ترکیه. ۲) از عوامل اصلی تشکیل و تداوم باد ۱۲۰ روزه سیستان و بطور کلی بادهای شمالی در شرق فلات ایران.

• نتایج بررسی ها نشان داد که ناوه‌ی شبه‌ساکن شرق ترکیه بطور متوسط در دوره‌ی پنج روزه‌ی دوم تا دوره‌ی پنج روزه‌ی سوم ماه ژوئن (۶-۱۰ تا ۱۱-۱۵ ژوئن) در نیمه‌ی شرقی ترکیه ظاهر می‌شود. این ناوه در حول و حوش ۶ تا ۱۰ ژوئیه به حداکثر عمق خود رسیده و پس از دوره‌ی پنج روزه‌ی سوم ماه سپتامبر (۱۵-۱۱ سپتامبر) در شرق ترکیه ناپدید می‌شود. ناوه‌ی شرق ترکیه از طریق سازوکارهای زیر بر اقلیم تابستانه ایران تأثیرگذار است: ۱) عمیق شدن ناوه‌ی شرق ترکیه بر نیمه‌ی غربی ایران و شکل گیری ناوه ای باریک و کشیده تا شمال خلیج فارس موجب تقویت و گسترش کم فشار ایران در طول تابستان می گردد. ۲) استقرار ناوه شبه ساکن امکان صعود دائمی هوا را در منطقه‌ای حدفاصل ۴۵° تا ۵۵° طول شرقی در تمام طول تابستان فراهم می‌آورد. بررسی ها نشان داد که میزان صعود هوا در امتداد طول ۵۰° شرقی به حداکثر میزان خود می رسد.

• نتایج بررسی ها نشان داد که کم فشار پاکستان در پایان دهه دوم ماه می در نتیجه‌ی افزایش تاوایی مثبت و پیدایش یک مرکز گردش چرخندی در طول ۷۰° شرقی تشکیل می شود و در طول فصل تابستان بطور متوسط در حدفاصل ۶۵° تا ۶۷° طول شرقی و ۳۰° عرض شمالی استقرار می‌یابد و در نهایت در دهه‌ی آخر ماه سپتامبر محو می‌گردد. کم فشار پاکستان بیشینه تاوایی مثبت خود را در تراز ۸۵۰ هکتوپاسکال تجربه نموده و گردش چرخندی آن حداکثر تا تراز ۷۰۰ هکتوپاسکال گسترش می یابد.

• بررسی‌های انجام شده نشان داد که در حول و حوش دوره‌ی پنج روزه‌ی اول ماه ژوئن زبانه‌ی کم‌فشاری بر روی غرب ایران و بر روی زاگرس شکل می‌گیرد. این زبانه کم فشار در طول فصل تابستان

جهتی شمال غربی-جنوب شرقی پیدا نموده و تا شرق مدیترانه گسترش می‌یابد. زبانه ی مذکور حداکثر گردش چرخندی خود را بر روی زاگرس در تراز ۸۵۰ هکتوپاسکال تجربه نموده، در ماه ژوئن به حداکثر گردش چرخندی خود رسیده و در دهه پایانی ماه سپتامبر محو می گردد. محاسبه گرمایش دررو با استفاده از معادله انرژی ترمودینامیک نشان داد که فرارفت قائم گرما از روی رشته کوه زاگرس عامل اصلی تشکیل و تداوم این زبانه‌ی کم‌فشار بر روی نیمه‌ی غربی ایران است. با توجه به موقعیت جغرافیایی و نحوه استقرار زبانه ی کم فشار، همچنین سازوکار حاکم بر آن، نام «زبانه ی کم فشار زاگرس» یا «ناوه ی فشاری زاگرس» برای آن پیشنهاد می شود. این ناوه فشاری از عوامل اصلی شکل گیری و تداوم باد شمال در نیمه غربی خلیج فارس و بستر اولیه تشکیل کم فشار ایران در طول تابستان است.

•بررسی جریان هوا در ترازهای زیرین جو بر روی جنوب غرب آسیا بیانگر ارتباط بین بادهای منطقه ای موسوم به «باد ۱۲۰ روزه سیستان» و «باد شمال» با نحوه استقرار مراکز فشار در وردسپهر زیرین است. با توجه به بررسی های انجام شده، باد۱۲۰ روزه سیستان نتیجه شیب فشار ناشی از شکل گیری و تداوم مرکز کم فشار پاکستان و واچرخند ترکمنستان به ترتیب در شرق و شمال شرق فلات ایران می باشد. در مقابل، وزش باد شمال در منطقه خلیج فارس نتیجه شکل گیری زبانه ی کم فشار زاگرس بر نیمه غربی ایران و جابجایی شمال غرب سوی مرکز پرفشار عربستان و استقرار مداوم آن بر بالای عرض ۲۷/۵° شمالی در طول فصل تابستان است.